\documentclass{manuscript}

\begin{document}

\preprint{APS/123-QED}

\title{Thermodynamically consistent coarse-graining of polar active fluids}

\author{Scott Weady$^1$}
\email{scott.weady@nyu.edu}
\author{David B. Stein$^2$}
\author{Michael J. Shelley$^{1,2}$}

\affiliation{$^1$Courant Institute of Mathematical Sciences, New York University, New York, NY, 10012, USA  \\ $^2$Center for Computational Biology, Flatiron Institute, New York, NY, 10010, USA}

\date{\today}
\begin{abstract}
We introduce a closure model for coarse-grained kinetic theories of polar active fluids. Based on a quasi-equilibrium approximation of the particle distribution function, the model closely captures important analytical properties of the kinetic theory, including its linear stability and the balance of conformational entropy production and dissipation. Nonlinear simulations show the model reproduces the qualitative behavior and nonequilibrium statistics of the kinetic theory, unlike commonly used closure models. We use the closure model to simulate highly turbulent suspensions in both two and three dimensions in which we observe complex multiscale dynamics, including large concentration fluctuations and a proliferation of polar and nematic defects.
\end{abstract}

\maketitle

\section{Introduction}

In many physical systems, microscopic structure dictates macroscopic dynamics. This is especially true in active fluids, where collections of work-performing particles and the hydrodynamic interactions between them can give rise to large-scale correlations and flows \cite{Ramaswamy:2010,Marchetti:2013,SS:2013,Bar:2020,Alert:2022}. Active fluids arise in a variety of physical contexts, with examples including suspensions of motile bacteria \cite{Dombrowski:2004,Sokolov:2007,Zhou:2014}, assemblies of microtubules cross-linked by molecular motors \cite{Koenderink:2009,Sanchez:2012}, and collections of liquid droplets whose interiors support chemical reactions \cite{Maass:2016,Gao:2017b,Young:2021}. 

Two main classes of active fluids are polar fluids and active nematics. Polar fluids are characterized by head-tail asymmetry in their constitutive particles, which may arise from propulsive mechanisms \cite{Pedley:1992}, polarity sorting \cite{GBGBS:2015,Furthauer:2019}, or asymmetric geometries \cite{Yamada:2003}. This asymmetry establishes a well-defined orientation, or polarity, of the particles in the fluid. On the other hand, nematic fluids, either active or passive, are classically composed of immotile, head-tail symmetric particles, which means particle orientation is invariant under sign change.

Though these two classes of systems have much in common, their microscopic differences produce unique macroscopic behaviors. For active nematics in two dimensions, characteristic features are $\pm 1/2$ topological defects \cite{Giomi:2013,Thampi:2016,Thampi:2013,Thampi:2014,Thijssen:2020}, which are points where the director field rotates $180^\circ$ clockwise or counter-clockwise around a point of zero nematic order, respectively. Three-dimensional analogs of $\pm1/2$ defects correspond to disclination lines and loops along which the director undergoes a variety of three-dimensional rotations \cite{Duclos:2020}. In polar fluids, orientation is typically described by a signed vector quantity called the polarity vector, though the director is still defined for such systems. The introduction of a well-defined sign to the orientation means the polarity vector does not exhibit $\pm1/2$ defects, but rather $\pm 1$ defects \cite{Elgeti:2011,Vafa:2020,Amiri:2022}. Polar fluids also demonstrate a range of flows which do not occur in active nematics, especially involving concentration instabilities and travelling waves driven by polar fluxes \cite{Voituriez:2006,SS:2008,Tjhung:2011,Giomi:2012}. 

Models of active fluids, either discrete or continuous, provide insight into the collective dynamics and its relationship to the system's microscopic structure. In discrete models, microscopic motion is explicitly simulated, with the hydrodynamic equations solved subject to the particles' self-generated flows \cite{Durlofsky:1987,SS:2007,Yan:2020}. Such models provide detailed insight into the multiscale dynamics, but are difficult to analyze and expensive to simulate when the number of particles is large. Continuum models, on the other hand, are formulated in terms of mean-field quantities that couple to the fluid through an active stress \cite{SS:2013}. Evolution equations for these mean-field quantities can be derived phenomenologically, for example by the variation of a free energy functional \cite{Liverpool:2008,Giomi:2012}, however this approach can obscure interpretation of the system parameters and their relationship to the underlying physics. 

Continuum kinetic theories are a specific class of models that are rooted in microscopic modeling \cite{Doi:1986,SS:2008b,ESS:2013}. In these models, the particle suspension is represented by a continuous distribution function that characterizes the number of particles at each point in space with a given orientation. The evolution of this distribution function can then be derived from conservation principles and explicit representations for the dynamics of the individual particles. However, since the distribution function depends on both position and orientation, these models are high dimensional and can be challenging to simulate. Coarse-graining the orientational degrees of freedom can mitigate this cost, where the system is instead represented by low-order orientational moments of the particle distribution function \cite{Woodhouse:2012,Gao:2017b,Chen:2018,Theillard:2019}. This yields evolution equations for mean-field quantities, like the polarity vector, that are similar to phenomenological models but have clearer physical origins. Unfortunately this does not come for free; in most cases the evolution equations for the orientational moments depend on unrepresented higher-order moments that must be approximated with a closure model. Various closure models have been proposed for coarse-grained kinetic theories, especially in the context of apolar suspensions \cite{Cintra:1995,FCL:1998,Baskaran:2008b,SS:2013}. Most of these are based on approximations that apply to specific flow regimes and, as a consequence, can produce unphysical solutions when applied ad-hoc \cite{FCL:1998}. 

In this paper, we propose a closure model for coarse-grained kinetic theories of polar active fluids. Based on a thermodynamically consistent, quasi-equilibrium approximation of the particle distribution function, the model generalizes the well-studied Bingham closure \cite{CL:1998,Han:2015,Li:2015,GBJS:2017,WSS:2022a} to account for polarity in microscopic structure. We first describe a continuum kinetic model for a concentrated polar active suspension and derive evolution equations for the first three orientational moments of the distribution function. We then motivate and introduce the closure model and show it closely approximates essential analytical properties of the kinetic theory, including the linearized behavior of the isotropic, polar aligned, and nematically aligned base states, and the evolution of the conformational entropy. These results are contrasted against first-moment models, which we find cannot capture the linear instability of the isotropic and nematic base states. Extending from our previous work on apolar closures \cite{WSS:2022a}, we outline a computational method for accurately and efficiently computing the closure in both two and three spatial dimensions. We use this method to study the nonlinear dynamics of the closure model and validate it against the kinetic theory, where we find good agreement in both the transient dynamics and nonequilibrium statistics of the system. In comparison, we find other closure models produce drastically different flows and inconsistent statistics, which in turn can lead to numerical instability. Finally, we demonstrate the capabilities of the model through high-resolution two- and three-dimensional simulations.

\section{The Doi-Saintillan-Shelley kinetic theory}

Here we summarize a continuum kinetic model for a concentrated  suspension of motile particles, further details can be found in References \cite{ESS:2013} and \cite{Subramanian:2009}. Consider a suspension of $N$ motile rod-like particles, each of length $\ell$ and diameter $b$, with aspect ratio $r = \ell/b \gg 1$. The particles are suspended in a fluid of linear dimension $L$ and volume $V = L^3$, from which we define a mean number density $\nu = N/V$. We describe the suspension by means of a continuous distribution function $\Psi(\x,\p,t)$ with $\int_V\int_{|\p|=1} \Psi ~ d\p d\x = N$, which characterizes the number of particles with center of mass $\x$ and orientation $\p$. Because the number of particles is conserved, this distribution function satisfies a Smoluchowski equation,
\begin{equation} \frac{\partial\Psi}{\partial t} + \grad_x\cdot(\dot\x\Psi) + \grad_p\cdot(\dot\p\Psi) = 0,\label{eq:dPsi/dt}\end{equation}
where $\grad_x$ is the spatial gradient and $\grad_p = (\I - \p\p)\cdot\partial_\p$ is the surface gradient on the unit sphere. The conformational flux functions $\dot\x$ and $\dot\p$ describe the translational and rotational dynamics of an individual particle, respectively, and depend on the specific microscopic model. For the model we consider, these fluxes are given by
\begin{align}
\dot\x &= V_0\p + \u - D_T\grad_x\log\Psi,\label{eq:xdot}\\
\dot\p &= (\I - \p\p)\cdot(\grad\u + 2\zeta_0c\Q)\cdot\p - D_R\grad_p\log\Psi,\label{eq:pdot}
\end{align}
where $\u(\x,t)$ is the fluid velocity, $(\grad\u)_{ij} = \partial u_i/\partial x_j$ is the velocity gradient, $c(\x,t) = \int_{|\p|=1} \Psi ~ d\p$ is the concentration, and $\Q(\x,t) = (1/c)\int_{|\p|=1} \p\p \Psi ~ d\p$ is the nematic tensor. Equation (\ref{eq:xdot}) says particles self-propel with velocity $V_0$, are advected with the local fluid velocity $\u$, and diffuse in space, where $D_T$ is the translational diffusion coefficient which, for simplicity, is assumed to be isotropic. Equation (\ref{eq:pdot}) describes contributions to rotational velocity arising from Jeffery's equation \cite{Jeffery:1922}, which says particles align with velocity gradients, and steric interactions modeled by the Maier-Saupe theory \cite{MS:1958}, which says particles tend to align with the principal axis of $\Q$, where $\zeta_0$ is the strength of steric alignment. The last term, proportional to $\grad_p\log\Psi$, models rotational diffusion, where $D_R$ is the rotational diffusion coefficient which is again assumed to be isotropic.

An essential feature in continuum models of suspensions, either active or passive, is the extra hydrodynamic stress due to the immersed particles. Here the stress has three components: the dipolar stress $\bSigma_a = \sigma_ac\Q$ generated by particle activity, a constraint stress due to particle rigidity $\bSigma_r = \sigma_rc\S:\E$, and stress due to steric interactions $\bSigma_s = -\sigma_sc^2(\Q\cdot\Q - \S:\Q)$, where $\S = (1/c)\int_{|\p|=1} \p\p\p\p\Psi ~ d\p$ is the fourth-moment tensor and $\E = (\grad\u + \grad\u^T)/2$ is the symmetric rate of strain tensor. The colon operator denotes contraction along the last two indices, that is, $(\S:\A)_{ij} = S_{ijk\ell}A_{k\ell}$. It can be shown that the dipole strength scales as $\sigma_a\propto\mu V_0\ell^2$, with a negative sign for pusher particles and a positive sign for puller particles. The coefficients $\sigma_r$ and $\sigma_s$ can be computed analytically, and are given by $\sigma_r = \pi\mu\ell^3/6\log(2r)$ and $\sigma_s = \pi\mu\ell^3\zeta_0/3\log(2r)$, where $\mu$ is the fluid viscosity. Finally, because the particle length and fluid velocity scales we typically consider are small, the total extra stress $\bSigma = \bSigma_a + \bSigma_r + \bSigma_s$ balances the Stokes equation
\begin{gather}
-\mu\Delta\u + \grad q = \div\bSigma,\label{eq:Stokes}\\
\div\u = 0,\label{eq:divu}
\end{gather}
where $q(\x,t)$ is the pressure. Equations (\ref{eq:dPsi/dt})-(\ref{eq:divu}) form a closed system which we refer to as the kinetic theory.

\subsection{Non-dimensionalization}

We choose a characteristic length scale $\ell_c = 1/\nu\ell^2$, velocity scale $u_c = V_0$, and time scale $t_c = \ell_c/u_c$, and normalize the distribution function by the number density $\nu$. The Smoluchowski equation (\ref{eq:dPsi/dt}) keeps the same form, and the particle flux functions (\ref{eq:xdot}) and (\ref{eq:pdot}) become
\begin{align}
\dot\x &= \p + \u - d_T\grad_x\log\Psi,\label{eq:xdot-nd}\\
\dot\p &= (\I - \p\p)\cdot(\grad\u + 2\zeta c\Q)\cdot\p - d_R \grad_p\log\Psi,\label{eq:pdot-nd}
\end{align}
where $d_T = (\nu\ell^2/V_0) D_T$ and $d_R = (1/\nu\ell^2 V_0) D_R$ are the dimensionless translational and rotational diffusion coefficients, respectively, and $\zeta = (1/\ell^2 V_0)\zeta_0$ is the dimensionless alignment strength. Under the same choice of characteristic scales, the Stokes equation becomes
\begin{gather}
-\Delta\u + \grad q = \div\bSigma,\label{eq:Stokes-nd}\\
\div\u = 0,\label{eq:divu-nd}
\end{gather}
with the dimensionless extra stress tensor
\begin{equation} \bSigma = \alpha c\Q + \beta c\S:\E - 2\beta\zeta c^2(\Q\cdot\Q - \S:\Q).\label{eq:stress-nd}\end{equation}
The parameter $\alpha = \sigma_a/\mu V_0\ell^2$ is the dimensionless dipole coefficient, which inherits the sign of $\sigma_a$, and $\beta = \pi \nu\ell^3/6\log(2r)$, being proportional to $\nu\ell^3$, characterizes the effective particle number density.

\subsection{Moment equations}\label{sec:coarse-grained}

The orientational degrees of freedom make the kinetic theory high dimensional which makes it expensive to simulate. To mitigate these difficulties, we can take orientational moments of the Smoluchowski equation (\ref{eq:dPsi/dt}) to represent the dynamics in terms of coarse-grained fields which depend only on space. To summarize, the relevant fields here are
\begin{align*}
 c(\x,t) &= \int_{|\p|=1} \Psi ~ d\p,\\
\n(\x,t) &= (1/c)\int_{|\p|=1} \p\Psi ~ d\p,\\
\Q(\x,t) &= (1/c)\int_{|\p|=1} \p\p\Psi ~ d\p,\\
\R(\x,t) &= (1/c)\int_{|\p|=1} \p\p\p\Psi ~ d\p,\\
\S(\x,t) &= (1/c)\int_{|\p|=1} \p\p\p\p\Psi ~ d\p,
\end{align*}
which are, respectively, the concentration $c$, the polarity vector $\n$, the nematic tensor $\Q$, and the third- and fourth-moment tensors $\R$ and $\S$. Integrating Eq. (\ref{eq:dPsi/dt}) against $1,\p$, and $\p\p$ over the unit sphere $\set{\p:|\p|=1}$ yields evolution equations for the first three moments $c,c\n$, and $c\Q$,
\begin{align}
\frac{Dc}{Dt} &= -\div(c\n) + H_c,\label{eq:dc/dt}\\
(c\n)^\grad  + c\R:\E &= - \div(c\Q) + \H_\n,\label{eq:dn/dt}\\
(c\Q)^\grad + 2c\S:\E  &= - \div(c\R) + \H_\Q\label{eq:dQ/dt},
\end{align}
where $D/Dt = \partial/\partial t + \u\cdot\grad$ is the material derivative, $\a^\grad = \partial\a/\partial t + (\u\cdot\grad)\a - \grad\u\cdot\a$ is the vectorial upper-convected time derivative, and $\A^\grad = \partial\A/\partial t + (\u\cdot\grad)\A - (\grad\u\cdot\A + \A\cdot\grad\u^T)$ is the tensorial upper-convected time derivative \cite{SS:2013}. The left hand side of the system of equations (\ref{eq:dc/dt})-(\ref{eq:dQ/dt}) consists of purely kinematic terms that are derived from hydrodynamic advection and rotation, where the contractions $(\R:\E)_i = R_{ijk}E_{jk}$ and $(\S:\E)_{ij} = S_{ijk\ell}E_{k\ell}$ enforce the constraint $|\p|=1$. The right hand side is separated into contributions from particle motility, captured by the polar fluxes $\div(c\n), \div(c\Q)$, and $\div(c\R)$, and terms arising from steric interactions, rotational diffusion, and translational diffusion
\begin{align}
    H_c &= d_T\Delta c,\label{eq:H_c}\\
    \H_\n &= 2\zeta c^2(\Q\cdot\n - \R:\Q) - (d-1)d_Rc\n + d_T\Delta(c\n),\label{eq:H_n}\\
    \H_\Q &= 4\zeta c^2(\Q\cdot\Q - \S:\Q) - 2dd_R\bpar{c\Q - \frac{c}{d}\I} + d_T\Delta(c\Q).\label{eq:H_Q}
\end{align}
The functions $H_c,\H_\n$, and $\H_\Q$ are analogous to those derived from the molecular free energy of Landau-deGennes-type theories \cite{deGennes:1993,BerisEdwards:1994}, but have a clear origin here in the microscopic dynamics. Note that the only source of polarity in Eqs. (\ref{eq:dc/dt})-(\ref{eq:dQ/dt}) is due to particle motility, which couples each moment to the next-order moment. In the absence of such terms, the equation for $c\n$ can be ignored, resulting in an apolar $Q$-tensor theory for a nematic suspension \cite{GBJS:2017}.

Formulated in terms of mean-field quantities, the moment equations (\ref{eq:dc/dt})-(\ref{eq:dQ/dt}) are similar to other models of polar fluids, such as those based on free-energy assumptions \cite{Liverpool:2008,Giomi:2012}, first-moment closures \cite{Reinken:2018}, or generalizations of the Toner-Tu theory \cite{Wensink:2012,Bratanov:2015}. Nearly all of these models neglect the second moment tensor equation (\ref{eq:dQ/dt}), effectively replacing the polar flux $\div(c\Q)$ by self-advection, $c(\n\cdot\grad)\n$, and the active stress $\alpha c\Q$ by the outer product $\alpha c\n\n$, both of which amount to approximating the nematic tensor as an outer product $\Q = \n\n$. This approximation is accurate under the assumption that particles are mostly polar aligned, that is $|\n|\approx 1$, however it cannot describe states that are nematically but not polar aligned. Moreover, it introduces additional nonlinearities in the first moment equation and the active stress, which leads to different instabilities and mechanisms of energy transfer across scales. By explicitly evolving the nematic tensor in addition to the polarity vector, the moment equations (\ref{eq:dc/dt})-(\ref{eq:dQ/dt}) show polarization can be generated from purely nematic states without nonlinear interactions.

\subsection{Closures of the moment equations}

As written above, the system of equations (\ref{eq:dc/dt})-(\ref{eq:dQ/dt}) is not closed because the third and fourth moments $\R$ and $\S$ are unrepresented. Evolution equations for these moments can be derived by integrating the Smoluchowski equation (\ref{eq:dPsi/dt}) against $\p\p\p$ and $\p\p\p\p$, respectively, which results in equations depending upon yet higher order moments and leads to the same issue. Instead, we choose to approximate these tensors in terms of the lower order moments $c,\n$, and $\Q$ with a closure model.

One of the most commonly used closure models is the isotropic closure \cite{Baskaran:2008,SS:2013,Theillard:2019}, which expresses the distribution function as a severely truncated series of spherical harmonics on the unit sphere,
\begin{equation} \Psi(\x,\p,t) = \frac{c}{2\pi(d-1)}\bpar{1 + d\n\cdot\p + a_d(\Q-\I/d):\p\p},\label{eq:iso-closure}\end{equation}
with constants $a_d = 4$ for $d = 2$ and $a_d = 15/2$ for $d = 3$, where $d$ is the spatial dimension. The moments $\R$ and $\S$ are then approximated as the third and fourth moments of (\ref{eq:iso-closure}), respectively. As suggested by its name, this model gives accurate solutions near isotropy $\Psi \approx 1/2\pi(d-1)$ and preserves the trace conditions $Q_{kk} = 1$, $R_{ikk} = n_i$, and $S_{ijkk} = Q_{ij}$. However, the isotropic closure performs poorly near aligned states where the distribution function has sharp orientational gradients and, in fact, the approximate distribution function (\ref{eq:iso-closure}) can take on negative values in regions of high polar alignment.

A different model that performs well when particles are nearly aligned is the quadratic closure \cite{FCL:1998}, which expresses the third and fourth moments as outer products,
\begin{align} 
R_{ijk} &= n_i Q_{jk},\label{eq:R-quad-closure}\\
S_{ijk\ell} &= Q_{ij}Q_{k\ell}\label{eq:S-quad-closure}.
\end{align}
This model also preserves the trace conditions on $\Q,\R$, and $\S$, but it introduces ad-hoc nonlinearities and is not based on a self-consistent representation of the distribution function. Moreover, the isotropic value for the fourth moment given by (\ref{eq:S-quad-closure}) is incorrect, which biases alignment. 

The Bingham closure is a self-consistent closure model that performs well for apolar suspensions, like active nematics, capturing the linearized behavior of the isotropic and nematically aligned states and preserving the balance of conformational entropy production and dissipation \cite{CL:1998,GBJS:2017,WSS:2022a}. In the apolar case, the distribution function is assumed to take the form of the Bingham distribution on the unit sphere \cite{Bingham:1974},
\begin{equation}\Psi_B(\x,\p,t) = Z(\x,t)^{-1} e^{\B(\x,t):\p\p},\label{eq:PsiB_apolar}\end{equation} 
where the symmetric matrix $\B$ and the normalization factor $Z$ are determined by constraints to match the zeroth and second moments $c$ and $\Q$ at each point in space. The resulting distribution function is then integrated against $\p\p\p\p$ to approximate the fourth-moment tensor $\S$. In this model odd moments are always zero, that is, $\n = 0$ and $\R = 0$, which restricts its applicability to apolar systems. 

\subsection{Quasi-equilibrium closure and the $B$-model}

The Bingham closure belongs to a more general class of closure models for kinetic equations, such as the Boltzmann equation, that are based on quasi-equilibrium approximation \cite{Levermore:1996,Levermore:1997,Abdelmalik:2016,Jiang:2021}. Interpreted in the context of suspensions, these models derive from the assumption that the orientational dynamics are fast, in which case the distribution function minimizes the conformational, or relative, entropy,
\begin{equation} \mathcal{S}[\Psi](t) = \int_V\int_{|\p|=1} \bpar{\frac{\Psi}{\Psi_0}}\log\bpar{\frac{\Psi}{\Psi_0}} ~ d\p d\x,\label{eq:entropy}\end{equation}
subject to pointwise constraints on known orientational moments, where $\Psi_0 = 1/2\pi(d-1)$ is the isotropic distribution function. (Note that this is defined with the opposite sign of the conventional information-theoretic entropy.) It's well known that such a distribution function takes the form of the so-called maximum-entropy distribution \cite{Cover:2012}, analogous to the Gibbs-Boltzmann distribution for a system in thermodynamic equilibrium,
\begin{equation} \Psi_B = Z^{-1} e^{-\beta_T\sum_n\mathcal{E}_n},\label{eq:Psi_GB}\end{equation}
where $Z$ is a normalization factor (the partition function) and $\beta_T$ is inverse temperature \cite{Callen:1985,Schroeder:1999}. The energies $\mathcal{E}_n = \B^{(n)}:\p^{(n)}$ here reflect the strength of alignment along a given axis, where $\B^{(n)}$ is a rank-$n$ tensor and $\p^{(n)}$ denotes the $n$th order outer product of $\p$. The moment equations can then be closed by solving for the parameters $\B^{(n)}$ based on the known moments, after which the distribution function (\ref{eq:Psi_GB}) can be integrated to obtain higher order moments. While this formulation is often used in classical kinetic theories, it has not yet been applied to polar suspensions, either active or passive.

Retaining up to the second moment, the distribution function (\ref{eq:Psi_GB}) is explicitly given by
\begin{equation} \Psi_B(\x,\p,t) = Z(\x,t)^{-1} e^{\B(\x,t):\p\p + \a(\x,t)\cdot\p},\label{eq:Psi_B}\end{equation}
where $\B$ is a traceless $d\times d$ symmetric matrix, $\a$ is a $d$-dimensional vector, and $Z$ is a scalar normalization factor. The parameters $\B$, $\a$, and $Z$ are determined by constraints to match the zeroth, first, and second moments at each point in space, 
\begin{align}
 c(\x,t) &= \int_{|\p|=1} \Psi_B ~ d\p,\label{eq:c-constraint}\\
\n(\x,t) &= (1/c)\int_{|\p|=1} \p\Psi_B ~ d\p,\label{eq:n-constraint}\\
\Q(\x,t) &= (1/c)\int_{|\p|=1} \p\p\Psi_B ~ d\p,\label{eq:Q-constraint}
\end{align}
after which the third- and fourth-moment tensors are approximated as
\begin{align}
\R_B[c,\n,\Q](\x,t) &= (1/c)\int_{|\p|=1} \p\p\p\Psi_B ~ d\p,\\
\S_B[c,\n,\Q](\x,t) &= (1/c)\int_{|\p|=1} \p\p\p\p\Psi_B ~ d\p.
\end{align}
For an apolar state, that is $\n = 0$, one can show the constraints (\ref{eq:c-constraint})-(\ref{eq:Q-constraint}) imply $\a = 0$ so that (\ref{eq:Psi_B}) reduces to the Bingham distribution. With this representation of $\R_B$ and $\S_B$, the first three moments evolve according to
\begin{align}
\frac{Dc}{Dt} &= -\div(c\n) + H_{c,B},\label{eq:dc/dt_b}\\
(c\n)^\grad  + c\R_B:\E &= - \div(c\Q) + \H_{\n,B},\label{eq:dn/dt_b}\\
(c\Q)^\grad + 2c\S_B:\E  &= - \div(c\R_B) + \H_{\Q,B}\label{eq:dQ/dt_b},
\end{align}
with
\begin{align}
    H_{c,B} &= d_T\Delta c,\label{eq:H_cB}\\
    \H_{\n,B} &= 2\zeta c^2(\Q\cdot\n - \R_B:\Q) - (d-1)d_Rc\n + d_T\Delta(c\n),\label{eq:H_nB}\\
    \H_{\Q,B} &= 4\zeta c^2(\Q\cdot\Q - \S_B:\Q) - 2dd_R\bpar{c\Q - \frac{c}{d}\I} + d_T\Delta(c\Q).\label{eq:H_QB}
\end{align}
Similarly, the stress tensor becomes
\begin{equation} \bSigma_B = \alpha c\Q + \beta c\S_B:\E - 2\beta\zeta c^2(\Q\cdot\Q - \S_B:\Q),\label{eq:stress-b}\end{equation}
and the Stokes equation
\begin{gather} -\Delta\u + \grad q = \div\bSigma_B,\\\div\u = 0,\label{eq:divu_b}\end{gather}
which is a closed system of equations in $c,\n$, and $\Q$. Throughout we refer to the mapping $(c,\n,\Q)\mapsto(\R_B,\S_B)$ along with the mean-field equations (\ref{eq:dc/dt_b})-(\ref{eq:divu_b}) as the $B$-model.

Despite its quasi-equilibrium formulation, the $B$-model has several important analytical properties that hold even out of equilibrium. First, it preserves the time evolution of the conformational entropy from the kinetic theory, satisfying the identity
\begin{equation} 
\begin{aligned}
\Psi_0 \frac{d\mathcal{S}[\Psi_B]}{dt} &= -\frac{2d}{\alpha}\int_V\E:\E ~ d\x - \int_V\int_{|\p|=1} (d_T|\grad_x\log\Psi_B|^2 + d_R|\grad_p\log\Psi_B|^2)\Psi_B ~ d\p d \x
\\ & =
\mathcal P(t) - \mathcal D(t),
\label{eq:dS/dt}
\end{aligned}
\end{equation}
which balances the rate-of-work with spatial and rotational dissipation. (For simplicity we've set $\beta = \zeta = 0$, however an equivalent identity holds for non-zero $\beta$ and $\zeta$; see \cite{GBJS:2017,WSS:2022a}.) A detailed proof of this identity for immotile suspensions is given in Ref. \cite{WSS:2022a}. The motile case follows from a similar argument so the proof is omitted here.  Further, by construction the trace conditions on $\Q,\R_B$ and $\S_B$ are automatically satisfied, as well as the feasibly attainable values of $c,\n,\Q,\R_B$, and $\S_B$. Finally, as we show in the next section, the $B$-model yields exact steady state solutions in the isotropic, polar aligned, and nematically aligned base states, and the linear theory of these states agrees well with the kinetic theory.

\section{Linear stability analysis}

Being formulated in terms of partial differential equations, continuum theories are useful tools for analyzing the hydrodynamic stability of particle suspensions \cite{Simha:2002,Voituriez:2006,ESS:2013}. Here we study analytically the linear stability of the $B$-model and compare with the kinetic theory. For simplicity we limit our discussion to three dimensions, finding analogous results in two dimensions.

\subsection{Isotropic base state}\label{sec:iso-stability}

We first consider the linear stability of the isotropic state $\Psi_0 \equiv 1/4\pi$ in a dilute suspension $\beta = \zeta = 0$ with zero diffusion $d_T = d_R = 0$. Taking $\B_0 = 0, \a_0 = 0$, and $Z_0 = 4\pi$ shows this is a valid solution of the $B$-model, with the corresponding mean-field quantities $c_0 = 1$, $\n_0 = 0$, and $\Q_0 = \I/3$. We consider perturbations in the first and second moments $\n = \n_0 + \n'$, $\Q = \Q_0 + \Q'$, and take $c = c_0$ as motivated by analysis of the kinetic theory \cite{SS:2008}. Writing $\Psi_B = \Psi_0\exp\{\lambda' + (\B_0+\B'):\p\p + (\a_0 + \a')\cdot\p\}$ and linearizing gives
\begin{equation*}\Psi_B \approx \Psi_0 + (\lambda' + \B':\p\p + \a'\cdot\p)\Psi_0.\end{equation*}
(Note that this is exactly the assumed distribution function in the isotropic closure (\ref{eq:iso-closure}), meaning the stability analysis equally applies to that model.) From the moments of this linearized distribution function we can solve for $\lambda',\B'$, and $\a'$, after which we find $\R_B' = 3\S_0\cdot\n'$, where $(\S_0)_{ijk\ell} = (1/15)(\delta_{ij}\delta_{k\ell} + \delta_{ik}\delta_{j\ell} + \delta_{i\ell}\delta_{jk})$ is the isotropic fourth-moment tensor. Ignoring terms that are quadratic in the primed variables, we find the perturbation fields satisfy
\begin{gather*}
\frac{\partial \n'}{\partial t} +  \div\Q' = 0,\label{eq:dn'/dt-iso}\\
\frac{\partial\Q'}{\partial t} + 3\div(\S_0\cdot\n') - \frac{2}{5}\E' = 0,\label{eq:dQ'/dt-iso}
\end{gather*}
and
\begin{gather}
-\Delta \u' + \grad q' = \alpha \div\Q',\label{eq:u'-iso}\\
\div\u' = 0,\label{eq:divu'-iso}
\end{gather}
with the constraint $\div\n' = 0$ which comes from the assumption $c' = 0$. Now considering plane-wave perturbations in each variable $\n' = \tn e^{\sigma t + i\k\cdot\x}$, $\Q' = \tQ e^{\sigma t + i\k\cdot\x}$, $\u' = \tu e^{\sigma t + i\k\cdot\x}$, and $q' = \tilde q e^{\sigma t + i\k\cdot\x}$, we get the set of algebraic equations
\begin{gather}
\sigma\tn + i\k\cdot\tQ = 0,\label{eq:tn}\\
\sigma\tQ + 3i\k\cdot(\S_0\cdot\tn) - \frac{i}{5}(\k\tu + \tu\k)  = 0,\label{eq:tQ}
\end{gather}
with the velocity given explicitly by
\begin{equation}\tu = \frac{i\alpha}{k}(\I - \hat\k\hat\k)\cdot\tQ\cdot\hat\k = -\frac{\alpha\sigma}{k^2}\tn,\label{eq:tu-isotropic}\end{equation}
where the last equality holds from Eq. (\ref{eq:tn}) and the constraint $i\k\cdot\tn = 0$. 
After a bit of algebra, we find the growth rates are
\begin{equation} \sigma_\pm = -\frac{\alpha}{10} \pm \frac{1}{10}\bpar{\alpha^2 - 20k^2}^{1/2}.\label{eq:iso-dispersion}\end{equation}
From this we see that for $\alpha < 0$ both $\sigma_+$ and $\sigma_-$ will be positive, while for $\alpha > 0$ both will be negative, meaning pushers, or extensile particles, are always unstable and pullers, or contractile particles, are linearly stable, as predicted by the kinetic theory.

\begin{figure*}[t!]
\centering
\includegraphics[scale=0.5]{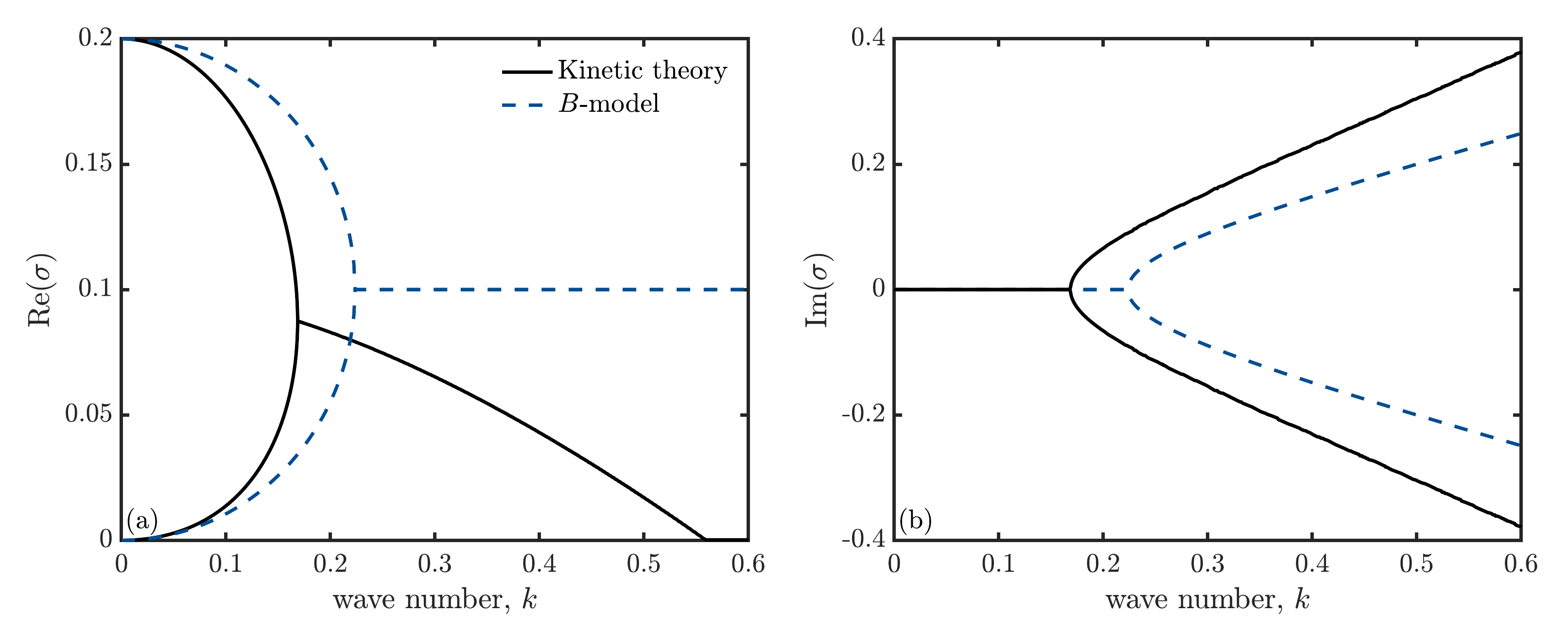}\vspace{-0.125in}
\caption{Dispersion relation $\sigma(k)$ for plane-wave perturbations about the isotropic base state $\Psi_0 = 1/4\pi$ for a dilute suspension of pusher particles with $\alpha = -1$ and $\beta = \zeta = 0$. The $B$-model is in close agreement with the kinetic theory at small wave numbers, capturing the bifurcation as well as the exact growth rate in the long wave limit $\sigma_+(k=0) = -\alpha/5$. For large wave numbers the growth rates diverge, with the $B$-model predicting all high wave numbers are unstable with a wave number independent growth rate $\sigma_{\pm} = -\alpha/10$ in the absence of spatial and rotational diffusion.}\label{fig:iso-dispersion}
\end{figure*}

Figure \ref{fig:iso-dispersion} shows the growth rates (\ref{eq:iso-dispersion}) along with those of the kinetic theory for $\alpha = -1$. In the long wave limit $k\rightarrow 0$ the dispersion relations are in increasingly better agreement, yielding the exact value for the $k = 0$ mode, which is always the most unstable. The $B$-model also captures the bifurcation, albeit at a slightly larger value $k = \sqrt{1/20} \approx 0.22$. A significant difference in the $B$-model is that, in the absence of diffusion, all wave numbers are unstable with a wave-number independent growth rate $\sigma_{\pm} = -\alpha/10$ beyond the bifurcation. We find this deviation from the kinetic theory is unavoidable with any pointwise closure model. In particular, the dispersion relation from the kinetic theory arises as the solution to a transcendental equation \cite{SS:2008}, while that of the coarse-grained theory is the solution of a constant coefficient eigenvalue problem and is therefore the root of a polynomial.

\begin{figure*}[t!]
\centering
\includegraphics[scale=0.5]{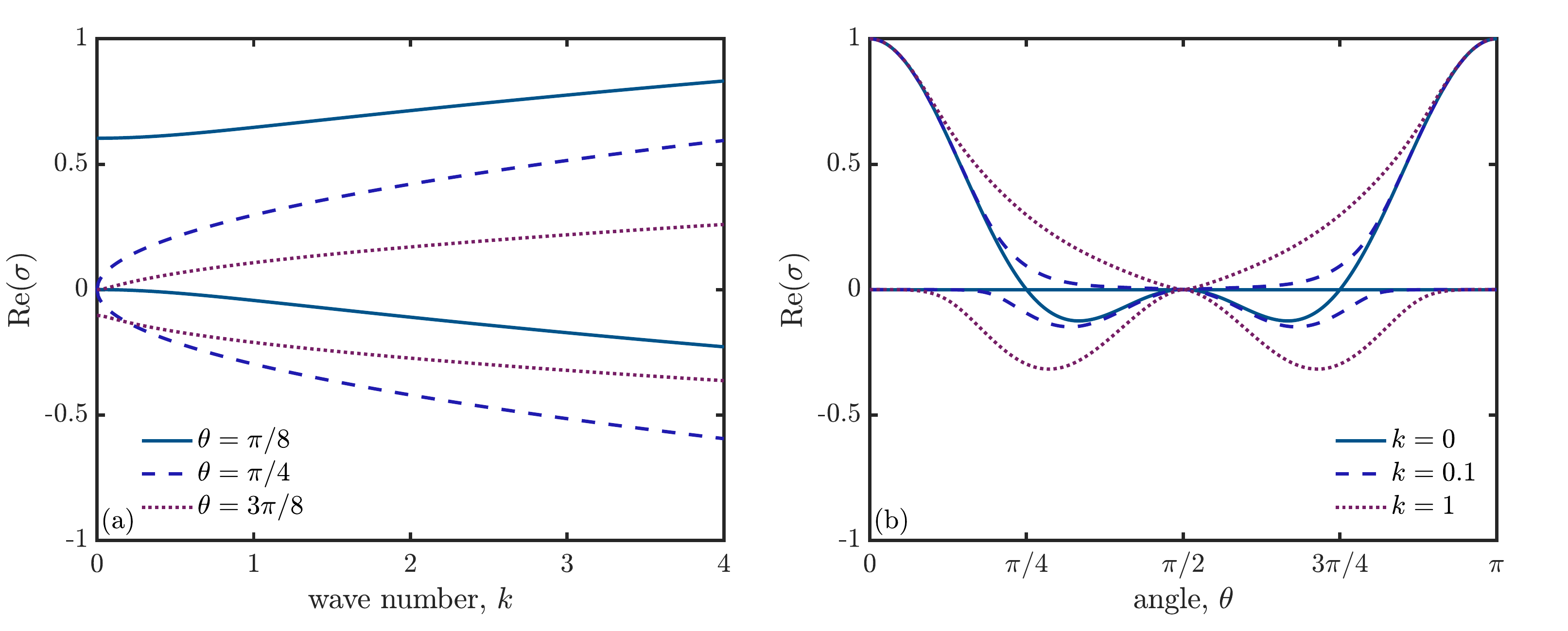}\vspace{-0.125in}
\caption{Dispersion relation for plane-wave perturbations about the sharply polar aligned base state $\lim_{\gamma\rightarrow\infty}Z^{-1}e^{\gamma\cos\theta}$ for a dilute suspension of pusher particles with $\alpha = -1$ and $\beta = \zeta = 0$. Growth rates are shown at several angles $\theta = \arccos(\hat\k\cdot\hat\z)$ and wave amplitudes $k$, where we find the base state is unstable in all cases. The $B$-model exactly agrees with the kinetic theory in this case, so only one dispersion relation is displayed.}\label{fig:aligned-dispersion}
\end{figure*}

This analysis also provides insight into the instability that is not as easily inferred from the kinetic theory. Namely, from Eq. (\ref{eq:tu-isotropic}) we see the perturbed polarity vector $\n'$ and the perturbed velocity $\u'$ are colinear, being parallel for pushers ($\alpha < 0$) and anti-parallel for pullers ($\alpha > 0$). This suggests polarity is positively correlated to velocity in pusher suspensions, a phenomenon which is not obvious \textit{a priori} but has been observed in numerical simulations \cite{SS:2008}. This also shows that the second-moment $\Q$ is necessary for capturing the linear instability. In particular, if only $c$ and $\n$ are evolved and we express $\Q := \Q[c,\n]$ and $\R := \R[c,\n]$, Eq. (\ref{eq:dn'/dt-iso}) for the perturbed polarity is closed and independent of the fluid velocity, which predicts linear stability as has been reported in previous studies of first-moment models \cite{Giomi:2012}.

\subsection{Polar aligned base state}

We next analyze the stability of polar aligned suspensions where all particles swim in the same direction, again considering dilute suspensions $\beta = \zeta = 0$ and neglecting diffusion $d_T = d_R = 0$. Here we consider base states of the form
\begin{equation}\Psi_\gamma(\theta) = Z^{-1} e^{\gamma\cos\theta},\label{eq:aligned-base-state}\end{equation}
where $\theta$ is the angle from the $z$-axis, and $\gamma$ is a constant which sets the strength of alignment. Taking $\B_0 = 0$ and $\a_0 = (0,0,\gamma\cos\theta)$ shows this distribution function is also of the form (\ref{eq:Psi_B}), and therefore corresponds to a valid solution of the $B$-model. We consider perturbations in each moment $c = 1 + c'$, $\n = \n_0 + \n'$, and $\Q = \Q_0 + \Q'$, where $\n_0 = \int_{|\p|=1}\p\Psi_\gamma ~ d\p$ and $\Q_0 = \int_{|\p|=1}\p\p\Psi_\gamma~d\p$ are the steady state polarity vector and nematic tensor, respectively.
%
%
Writing $\Psi_B = \Psi_\gamma\exp\{\lambda' + \B':\p\p + \a'\cdot\p\}$ and linearizing, we can derive
\begin{align}
\n' &=  \n_0\lambda' + \R_0:\B' + \Q_0\cdot\a',\label{eq:n'}\\
\Q' &=  \Q_0\lambda' + \S_0:\B' + \R_0\cdot\a',\label{eq:Q'}\\
\R_B' &=  \R_0\lambda' + \T_0:\B' + \S_0\cdot\a',\label{eq:R'}
\end{align} 
where $\R_0 = \int_{|\p|=1}\p\p\p\Psi_\gamma ~ d\p$, $\S_0 = \int_{|\p|=1} \p\p\p\p\Psi_\gamma ~ d\p$, and $\T_0 = \int_{|\p|=1}\p\p\p\p\p \Psi_\gamma ~ d\p$. The quantities $\n_0$, $\Q_0$, $\R_0$, $\S_0$, and $\T_0$ can be computed analytically, after which we take the sharply aligned limit $\gamma\rightarrow\infty$ to get $\n_0 = \hat\z$, $\Q_0 = \hat\z\hat\z$, $\R_0 = \hat\z\hat\z\hat\z$, $\S_0 = \hat\z\hat\z\hat\z\hat\z$, and $\T_0 = \hat\z\hat\z\hat\z\hat\z\hat\z$. Equations (\ref{eq:n'})-(\ref{eq:R'}) then imply $\Q' = \hat\z\n' + \n'\hat\z$ and $\R' = \hat\z\hat\z\n' + \hat\z\n'\hat\z + \n'\hat\z\hat\z$ which, neglecting terms that are quadratic in the primed variables, gives a closed system of equations for the perturbations $c'$ and $\n'$,
\begin{align*}
&\frac{\partial c'}{\partial t} + \hat\z\cdot\grad c' + \div\n' = 0,\\
&  \frac{\partial \n'}{\partial t} + \hat\z\cdot\grad\n' - (\I - \hat\z\hat\z)\cdot\grad\u'\cdot\hat\z = 0,
\end{align*}
with
\begin{gather*}
-\Delta\u' + \grad q' = \alpha\div(c'\hat\z\hat\z + \hat\z\n' + \n'\hat\z),\\
\div\u' = 0.
\end{gather*}
This system is identical to that derived in \cite{SS:2008}, for which the dispersion relation was found to be
\begin{equation} \sigma_\pm = \frac{1}{2}f(\theta)\cos2\theta\bbrack{1 \pm \bpar{1 - \frac{4ik}{\alpha}\frac{\sin^2\theta\cos\theta}{f(\theta)\cos^22\theta}}^{1/2}} - ik\cos\theta,\label{eq:aligned-dispersion}\end{equation}
where $f(\theta) = -\alpha\cos^2\theta$ with $\cos\theta = \hat\k\cdot\hat\z$. This dispersion relation is plotted in Fig. \ref{fig:aligned-dispersion}, where we find for all angles at least one of $\text{Re}(\sigma_\pm)$ is positive, meaning the polar-aligned base state is unstable to all perturbations. Exact agreement with the kinetic theory for this base state suggests the $B$-model should be increasingly accurate when the particles are highly aligned. Note that this analysis is only valid for vanishing rotational diffusion, otherwise the suspension will always relax to an orientationally isotropic state. However, the analysis is still relevant on dimensionless timescales that are small relative to $1/d_R$ for which rotational diffusion can be neglected.

\subsection{Nematically aligned base state}

When steric effects and rotational diffusion are included, i.e. $\beta,\zeta,d_R > 0$, additional steady state solutions exist which are the so-called nematic base states,
\[ \Psi_\delta(\theta) = Z^{-1} e^{\delta \cos 2\theta},\]
where the exponent $\delta$ depends on the alignment parameter $\xi = 2\zeta/d_R$. These states are also of the form (\ref{eq:Psi_B}) and hence are valid solutions of the $B$-model. (Note that for such base states the polarity vector vanishes, meaning they cannot be described by a first-moment closure model.) Using the Bingham closure, Gao et al. \cite{GBJS:2017} analyzed the linear stability of the nematic base states for immotile suspensions and found good agreement between the closure model and the kinetic theory, especially at large values of $\xi$. Their analysis equally holds for the $B$-model by neglecting the swimming contributions and taking $\a \equiv 0$. When steric interactions are strong, particle motility has a relatively minor influence on stability \cite{ESS:2013}, so we reserve a more detailed analysis for future work.

\section{Outline of numerical implementation}

The $B$-model consists of an intermediate mapping from the moments $(c,\n,\Q)$ to the parameters $(Z,\a,\B)$, after which the distribution function $\Psi_B = Z^{-1}e^{\B:\p\p+\a\cdot\p}$ is integrated to obtain the third and fourth moment tensors $\R_B$ and $\S_B$. Starting from the constraints (\ref{eq:c-constraint})-(\ref{eq:Q-constraint}), we can solve for the normalization constant $Z$ to get
\begin{align}
\n &= \frac{\int_{|\p|=1} \p e^{\B:\p\p + \a\cdot\p} ~ d\p}{\int_{|\p|=1} e^{\B:\p\p + \a\cdot\p} ~ d\p},\label{eq:n-constraint-2}\\
\Q &= \frac{\int_{|\p|=1} \p\p e^{\B:\p\p + \a\cdot\p} ~ d\p}{\int_{|\p|=1} e^{\B:\p\p + \a\cdot\p} ~ d\p},\label{eq:Q-constraint-2}
\end{align}
which is independent of concentration. As written, this is a $(d^2/2 + 3d/2 - 1)$-dimensional nonlinear system (5 equations in 2D and 8 equations in 3D). Moreover, solving this nonlinear system requires evaluating many integrals on the unit sphere which is expensive to do at each spatial grid point and time step. Alternatively, we can precompute the closure map $(\n,\Q)\mapsto(\R_B,\S_B)$ and interpolate during simulations. This approach is particularly efficient near aligned states for which the system of equations (\ref{eq:n-constraint-2})-(\ref{eq:Q-constraint-2}) becomes ill-conditioned, requiring many quadrature nodes to evaluate the integrals and a large number of Newton iterations. Here we outline a framework for constructing Chebyshev interpolants for the closure map $(\n,\Q)\mapsto(\R_B,\S_B)$; further details can be found in Appendix \ref{app:interp} and in Reference \cite{WSS:2022a}. A MATLAB implementation is also available on GitHub \footnote{Implementations of the closure maps are available at \texttt{https://github.com/scott-weady/polar-closure}.}.

Because $\Q$ is symmetric, it has an eigendecomposition of the form $\Q = \bOmega\tQ\bOmega^T$, where $\tQ$ is a diagonal matrix consisting of the ordered eigenvalues of $\Q$, and $\bOmega$ is an orthonormal matrix whose columns are the ordered eigenvectors of $\Q$. Multiplying Eq. (\ref{eq:n-constraint-2}) by $\bOmega^T$ and conjugating Eq. (\ref{eq:Q-constraint-2}) by $\bOmega$ gives
\begin{align}
\tn &= \frac{\int_{|\tp|=1} \tp e^{\tB:\tp\tp + \ta\cdot\tp} ~ d\tp}{\int_{|\tp|=1} e^{\tB:\tp\tp + \ta\cdot\tp} ~ d\tp},\label{eq:tq-constraint}\\
\tQ &= \frac{\int_{|\tp|=1} \tp\tp e^{\tB:\tp\tp + \ta\cdot\tp} ~ d\tp}{\int_{|\tp|=1} e^{\tB:\tp\tp + \ta\cdot\tp} ~ d\tp},\label{eq:tD-constraint}
\end{align}
where $\tn = \bOmega^T\n$, $\ta = \bOmega^T\a$, and $\tB = \bOmega^T\B\bOmega$, and we've reparameterized the unit sphere by $\tp = \bOmega^T\p$. Though this nonlinear system is still $(d^2/2 + 3d/2 - 1)$-dimensional, the off-diagonal terms of $\tQ$ vanish and so the mapping $(\tn,\tQ)\mapsto(\tR_B,\tS_B)$ is ($2d-1)$-dimensional, where we've defined
\begin{align}
\tR_B &= \frac{\int_{|\tp|=1} \tp\tp\tp e^{\tB:\tp\tp + \ta\cdot\tp} ~ d\tp}{\int_{|\tp|=1} e^{\tB:\tp\tp + \ta\cdot\tp} ~ d\tp},\label{eq:tR}\\
\tS_B &= \frac{\int_{|\tp|=1} \tp\tp\tp\tp e^{\tB:\tp\tp + \ta\cdot\tp} ~ d\tp}{\int_{|\tp|=1} e^{\tB:\tp\tp + \ta\cdot\tp} ~ d\tp},\label{eq:tS}
\end{align}
as the rotated third- and fourth-moment tensors. 

In the rotated coordinate system the constraints on the feasible values of the moments $(\tn,\tQ)$ take a particularly simple form. Constraints on the first moment come from the variance inequalities $\langle p_i \rangle^2 \leq \langle p_i^2 \rangle$, yielding $\tilde n_i^2 \leq \tilde Q_{ii}$, for $i = 1,\ldots,d$. Additional constraints arise from the trace condition $\sum_{i=1}^d Q_{ii} = 1$, and the eigenvalue ordering $0\leq\tilde Q_{22} \leq \tilde Q_{11} \leq 1$ or $0 \leq \tilde Q_{33} \leq \tilde Q_{22} \leq \tilde Q_{11}\leq 1$. Establishing a mapping between this feasible domain and the hypercube $[-1,1]^{(2d-1)}$ (see Appendix) admits a Chebyshev representation for the closure map $(\tn,\tQ)\mapsto(\tR_B,\tS_B)$ that can be efficiently evaluated in practice.

While $\tR_B$ and $\tS_B$ have many components, we can reduce the number of interpolants to evaluate using symmetries and the trace identities
\begin{align}
\tilde n_i &= \tilde R_{ikk},\label{eq:trace(R)}\\
\tilde Q_{ij} &= \tilde S_{ijkk},\label{eq:trace(S)}
\end{align}
where repeated indices imply summation. Finally, after computing $\tR_B$ and $\tS_B$, we rotate back to the original frame using the identities
\begin{align}
R_{ijk} &= \Omega_{im}\Omega_{jn}\Omega_{kp} \tilde R_{mnp},\\
S_{ijk\ell} &= \Omega_{im}\Omega_{jn}\Omega_{kp}\Omega_{\ell q} \tilde S_{mnpq},
\end{align}
along with the analogous trace conditions (\ref{eq:trace(R)}) and (\ref{eq:trace(S)}) in the original coordinate system.

\begin{figure*}[t!]
\centering
\includegraphics[scale=0.5]{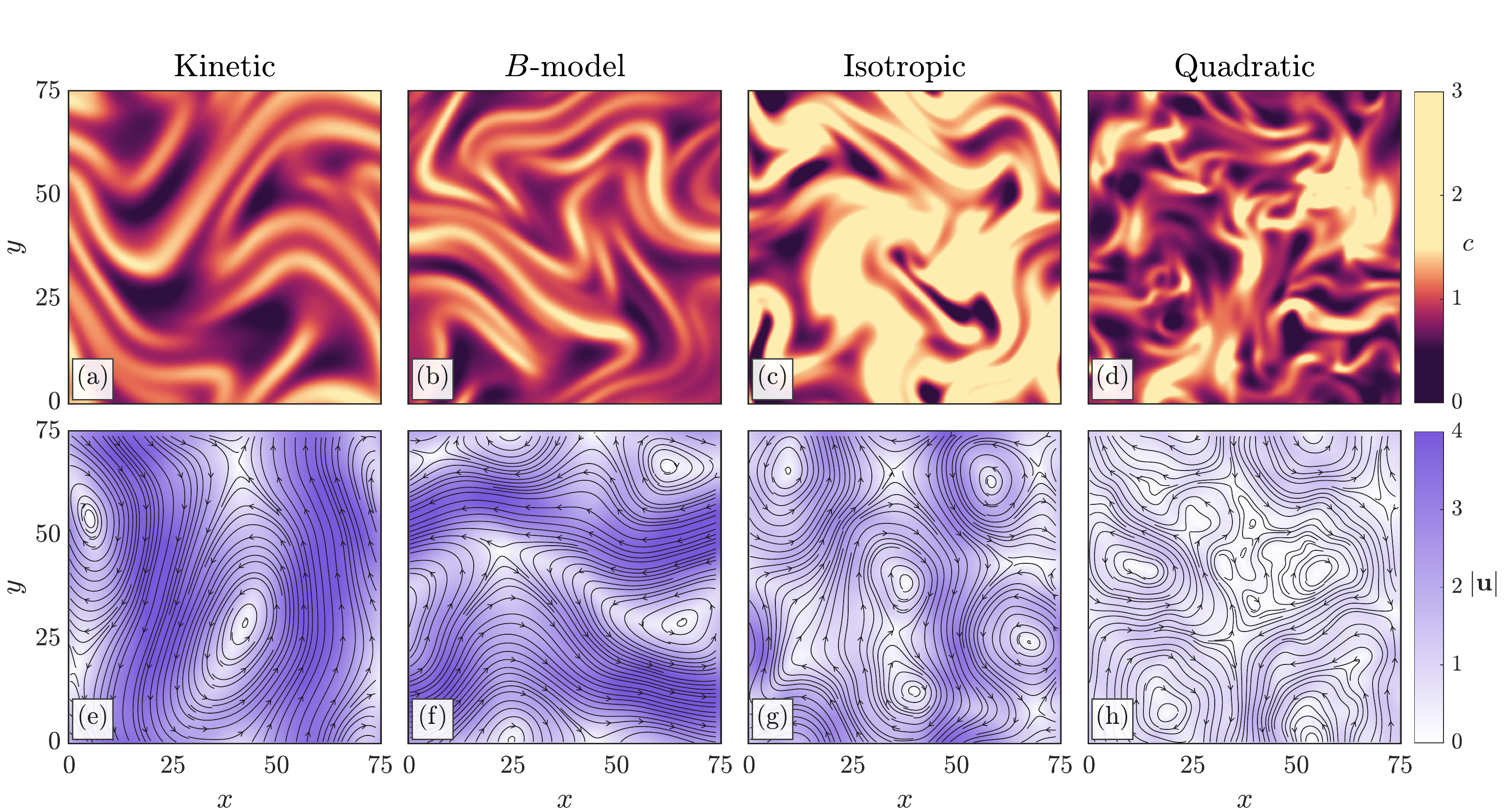}\vspace{-0.125in}
\caption{Comparison of computed concentration (a)-(d) and velocity (e)-(h) fields for the kinetic theory, the $B$-model, and the isotropic and quadratic closures for a dilute suspension with $\alpha = -1$ and $\beta = \zeta = 0$. The diffusion coefficients are $d_T = 0.2$ and $d_R = 0.02$, and the box size is $L = 75$. Qualitatively, the $B$-model is in much better agreement with the kinetic theory, while the isotropic and quadratic closures significantly differ in the length and velocity scales, as well as the magnitude of concentration fluctuations.}\label{fig:fields-dilute}
\end{figure*}

\section{Nonlinear dynamics and comparison with the kinetic theory}\label{sec:bingham-vs-kinetic}

To assess the long-time dynamics of the $B$-model, we perform two-dimensional numerical simulations in periodic geometries. For all simulations the discretization is pseudo-spectral using $256^2$ spatial Fourier modes and the 2/3 anti-aliasing rule, for which analysis of the Fourier spectrum shows the simulations are sufficiently resolved,  and time-stepping is done with a second order, implicit-explicit, backwards-differentiation scheme. For the kinetic theory we represent the orientation vector in polar coordinates $\p = (\cos\theta,\sin\theta)$, $\theta \in [0,2\pi)$, and discretize with $64$ orientational Fourier modes in $\theta$. For the $B$-model we use $M = 50$ degree Chebyshev interpolants for all closure maps. The initial data is a perturbation from isotropy $\Psi_0 = 1/2\pi$ using a distribution function of the form (\ref{eq:Psi_B}), and the simulation is run well past the initial instability to measure long-time statistics, with the initial conditions for the coarse-grained simulations computed as moments of the same initial distribution function. For ease of implementation, in the simulations we evolve the unnormalized moments $c$, $c\n$ and $c\Q$, which are linear in $\Psi$. We also implemented identical schemes for the linear and quadratic closure models.

\subsection{Dilute suspensions}\label{sec:dilute}

We first study the nonequilibrium dynamics of the $B$-model for a dilute suspension ($\beta = \zeta = 0$) so that any instabilities are solely caused by particle activity. For these simulations we consider pusher particles with dipole strength $\alpha = -1$, and translational and rotational diffusivities $d_T = 0.2$ and $d_R = 0.02$. The time step for all simulations in this section is $\Delta t = 0.01$.

Figure \ref{fig:fields-dilute} compares the concentration (a)-(d) and velocity fields (e)-(h) from simulations of the kinetic theory, the $B$-model, and the isotropic and quadratic closures at time $t = 200$, for which each model has reached a quasi-steady chaotic state. The overall scales of concentration and velocity fluctuations in the kinetic theory and the $B$-model are in good agreement, while both the linear and quadratic closures fail to produce solutions that qualitatively match the kinetic theory in either length or velocity scales. We found that for the isotropic closure to remain numerically stable for these parameters it was necessary to numerically enforce $c \geq 0$, which resulted in the growth of total concentration over time.

\begin{figure}[t!]
\centering
\includegraphics[scale=0.5]{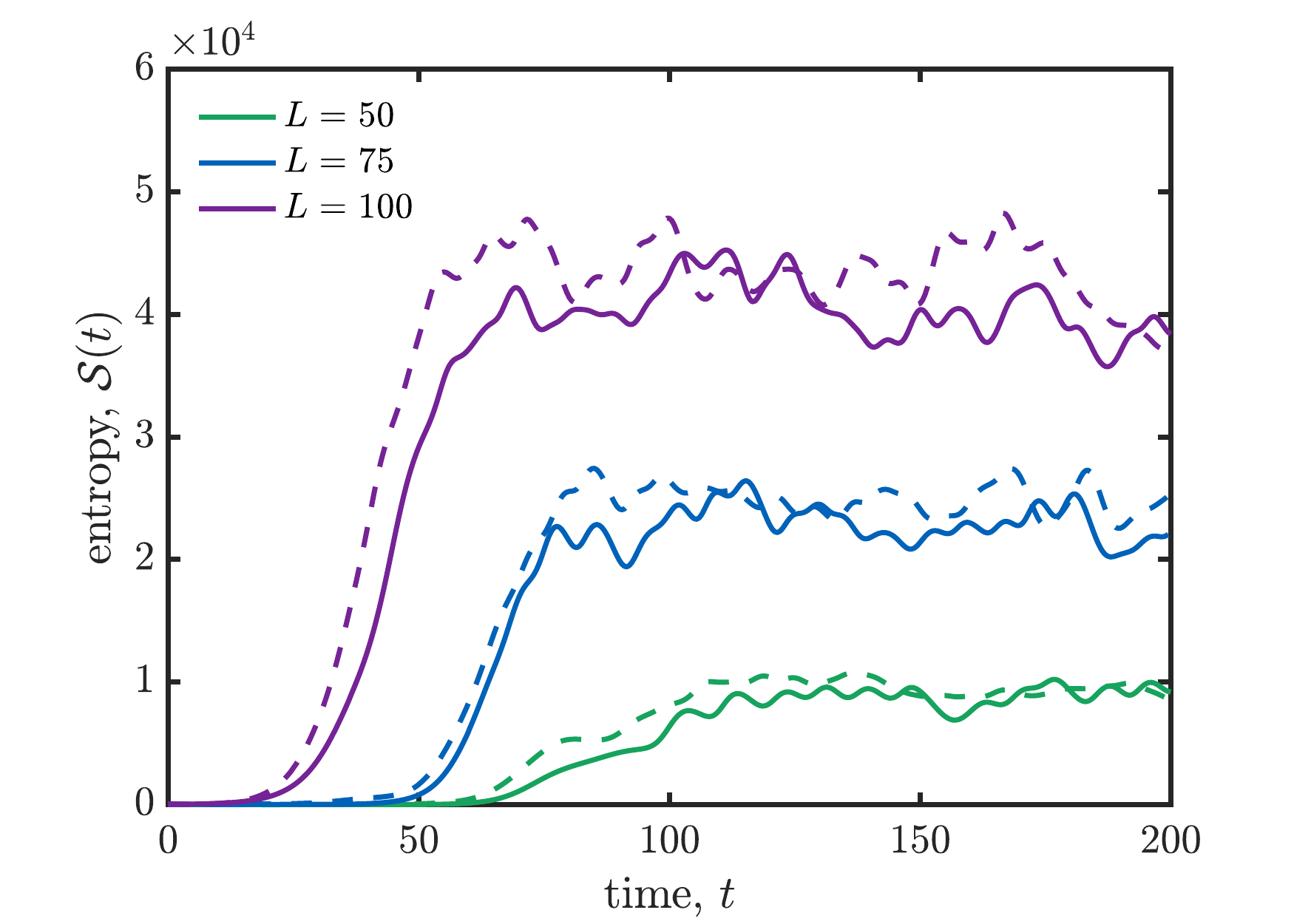}\vspace{-0.125in}
\caption{Evolution of the conformational entropy (\ref{eq:entropy}) for the kinetic theory (solid) and the $B$-model (dashed) in a dilute suspension of pushers for increasing box size $L$. The parameters are $\alpha = -1$ and $\beta = \zeta = 0$, with diffusion coefficients $d_T = 0.2$ and $d_R = 0.02$. For each case the $B$-model grows from isotropy slightly faster than the kinetic theory, which is consistent with linear stability analysis. At later times the $B$-model closely matches the mean entropy as well as fluctuations about the mean.}\label{fig:entropy-dilute}
\end{figure}

A useful property of the $B$-model is that the distribution function has an analytical, self-consistent form. This provides an approximation to statistics of the microscopic structure, such as the conformational entropy $\mathcal S(t)$, defined in Eq. (\ref{eq:entropy}). Figure \ref{fig:entropy-dilute} shows the evolution of $\mathcal S(t)$ for systems with box sizes $L = 50,75$, and $100$. As predicted by the linear stability analysis in Section \ref{sec:iso-stability}, the $B$-model grows away from isotropy slightly faster than the kinetic theory for all box sizes tested. The conformational entropy in both models eventually fluctuates about a quasi-steady mean, with the $B$-model yielding an accurate estimate of the mean as well as the magnitude of fluctuations.

\begin{figure}[t!]
\centering
\includegraphics[scale=0.5]{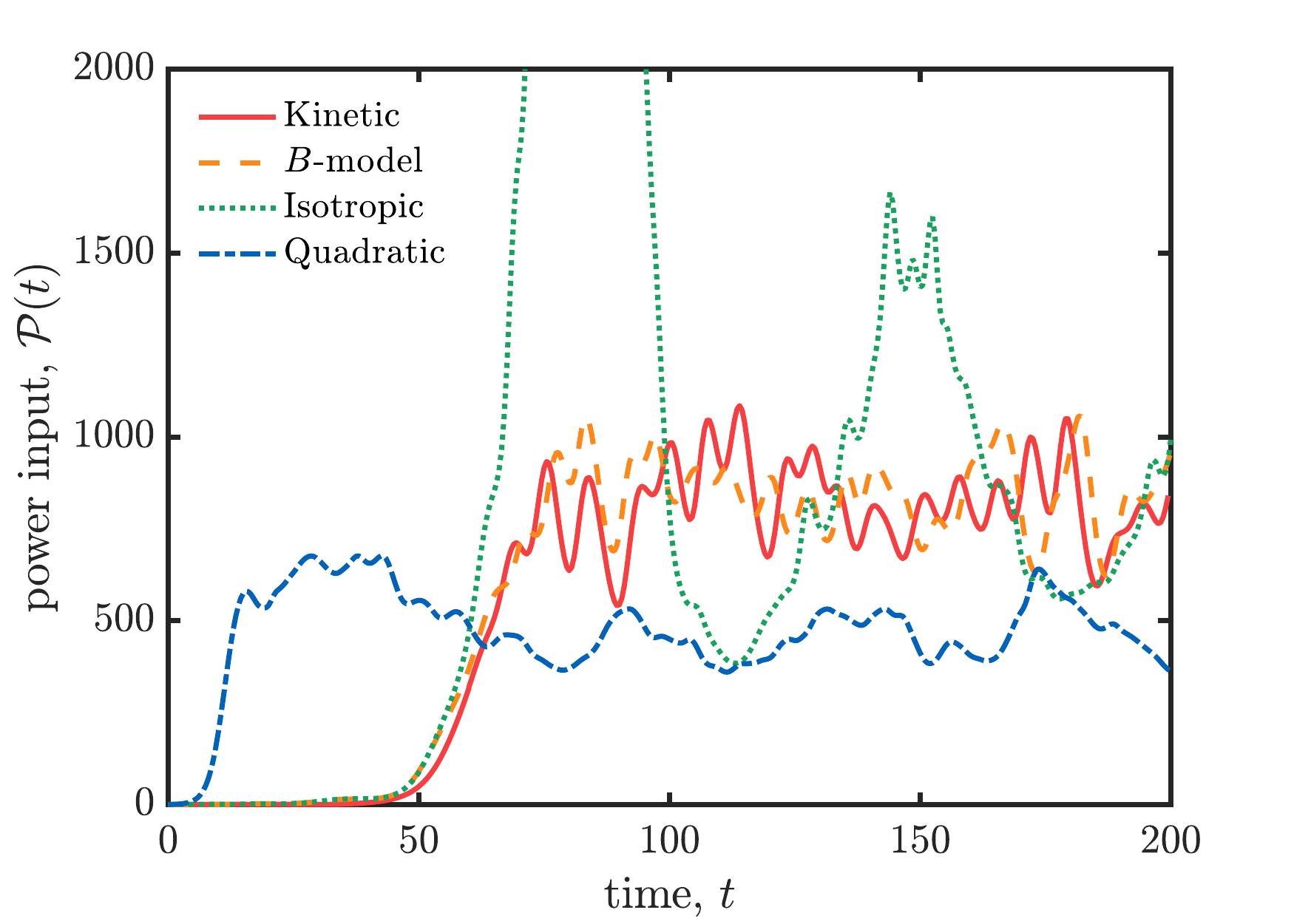}\vspace{-0.125in}
\caption{Evolution of the power input (\ref{eq:power}) for the kinetic theory, the $B$-model, and the isotropic and quadratic closures from the same simulations shown in Fig. \ref{fig:fields-dilute}.  The $B$-model accurately captures the mean input power, while the isotropic and quadratic closures differ significantly in both the mean and fluctuations about the mean.}\label{fig:power-dilute}
\end{figure}

It's also informative to compare the statistics of the $B$-model to other closure models. Here we consider, as an informative measure of activity, the input power 
\begin{equation}
\mathcal P(t) = -\frac{4}{\alpha}\int_V \E:\E ~ d\x,\label{eq:power}
\end{equation}
which, for extensile particles, $\alpha < 0$, is a source of conformational entropy according to Eq. (\ref{eq:dS/dt}). The parameters are the same as those in Figs. \ref{fig:fields-dilute} and \ref{fig:entropy-dilute}, with the box size fixed at $L = 75$. Figure \ref{fig:power-dilute} shows the evolution of $\mathcal P(t)$ for the kinetic theory along with the $B$-model and the isotropic and quadratic closures. The power input for the $B$-model is in close agreement with the kinetic theory, capturing comparable growth rates at early times as well as the long-time mean. The isotropic closure performs similarly at early times, showing the same growth rate as the $B$-model, which is to be expected from the linear stability analysis in Section \ref{sec:iso-stability}. At late times, however, the isotropic closure severely deviates from the kinetic theory, exhibiting large fluctuations without a clear mean. In contrast, the quadratic closure rapidly grows away from isotropy and underestimates the mean power input at late times by nearly a factor of two, although the magnitude of fluctuations is comparable. The faster growth rate of the quadratic closure is also consistent with expectations due to its inability to capture the isotropic state.

\subsection{Concentrated suspensions}

We next compare the nonequilibrium dynamics of the $B$-model for concentrated suspensions with steric interactions. Here we focus on similarities and differences between polar and nematic order, which is made accessible by evolving both the first and second moment tensors. We again consider pusher particles with dipole strength $\alpha = -1$ and particle density $\beta = 0.8$, and vary the alignment strength $\zeta$. As before, the diffusion coefficients are $d_T = 0.2$ and $d_R = 0.02$, which ensures both the isotropic and nematic base-states are linearly unstable \cite{ESS:2013}. The box size for all of these simulations is $L = 50$, and the time step is $\Delta t = 0.01$.

\begin{figure*}[t!]
\centering
\includegraphics[scale=0.5]{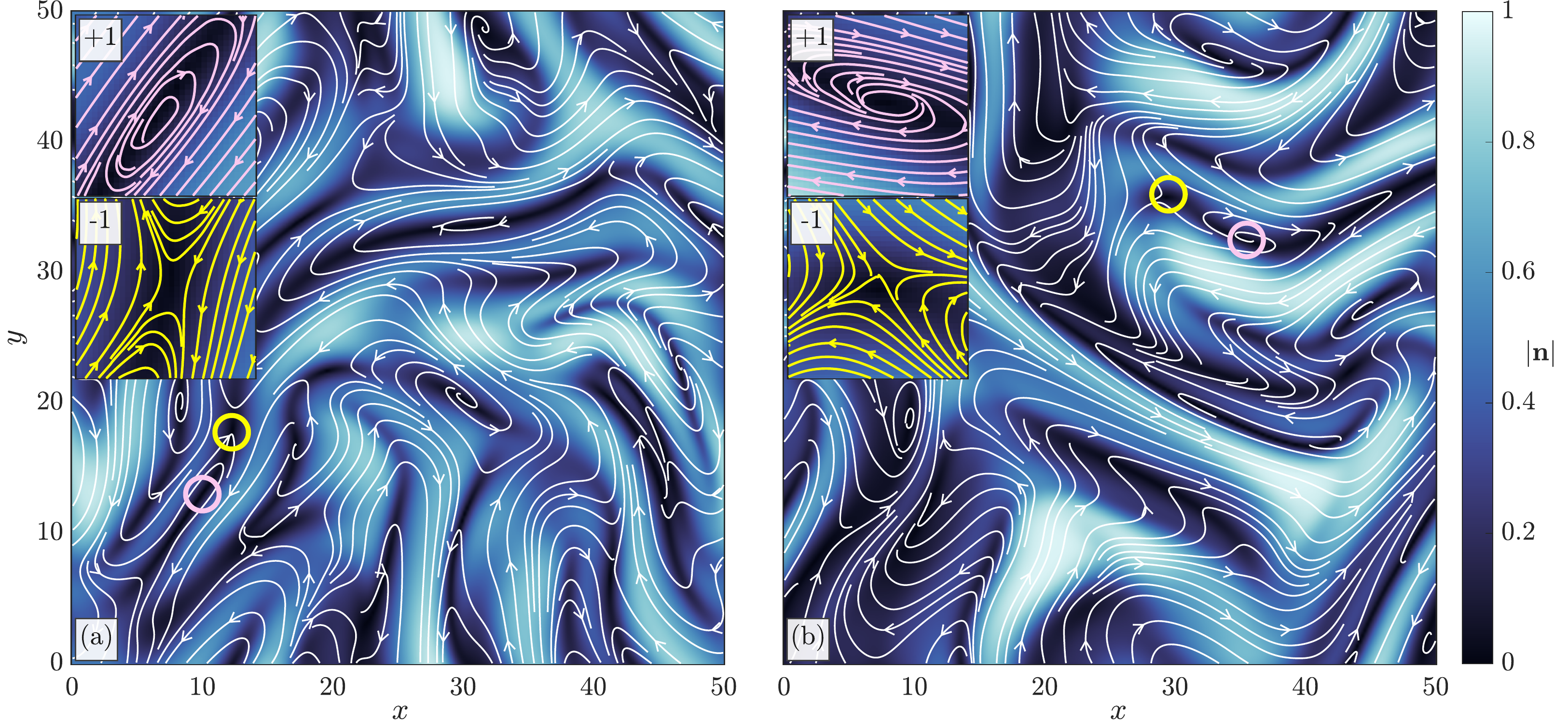}\vspace{-0.125in}
\caption{The scalar polar order parameter $n(\x,t) = |\n(\x,t)|$ and integral curves of the polarity vector field $\n$ in a concentrated suspension for the (a) kinetic theory and (b) $B$-model. The dimensionless parameters are $\alpha = -1$, $\beta = 0.8$, $\zeta = 0.2$, $d_T = 0.2$, and $d_R = 0.02$, and the box size is $L = 50$. The overall scales and features are qualitatively similar in both models, with the $B$-model accurately reproducing $\pm 1$ defect structures, shown in the insets and indicated by pink and yellow circles, respectively.}\label{fig:polar-order-2d}
\end{figure*}

\begin{figure*}[t!]
\centering
\includegraphics[scale=0.5]{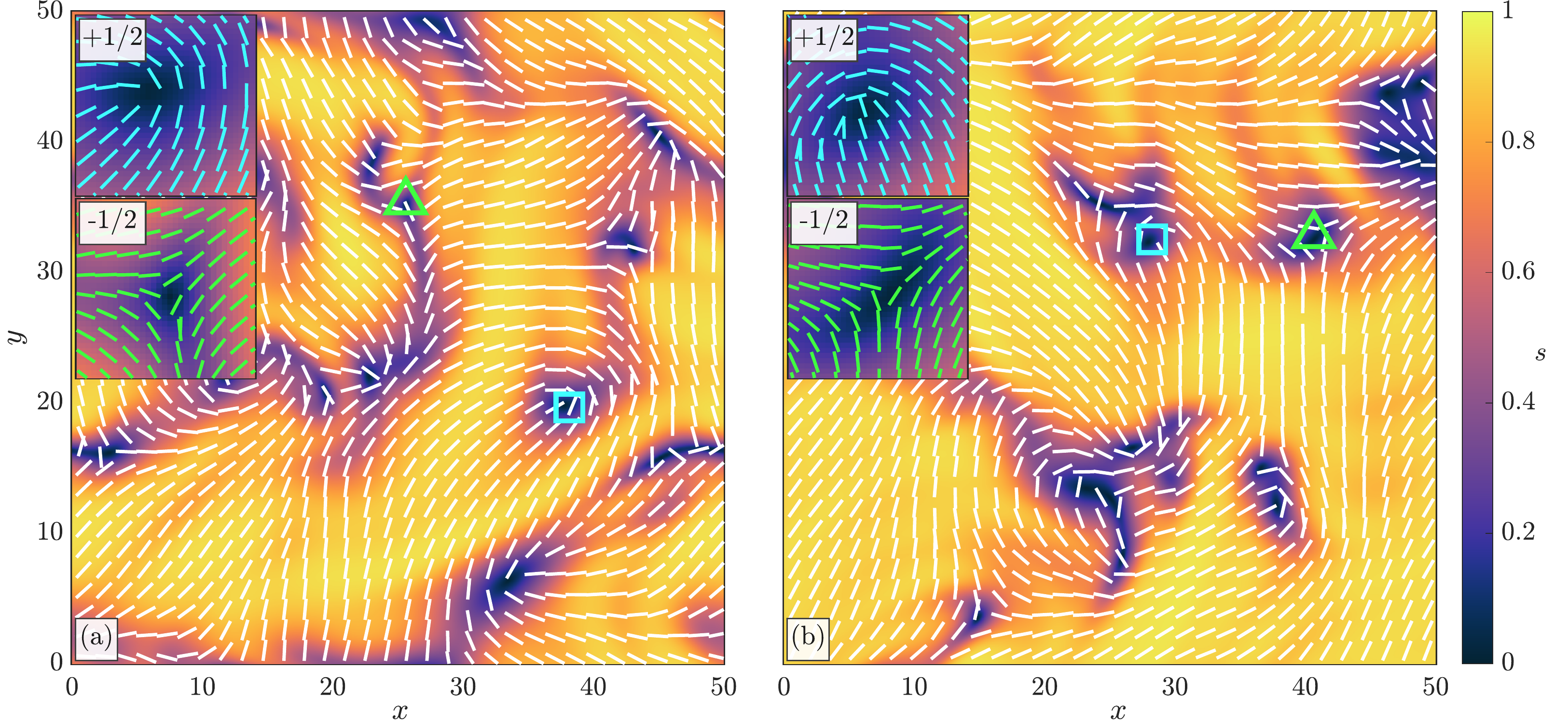}\vspace{-0.125in}
\caption{The scalar nematic order parameter $s(\x,t)$ and director field $\m$ (white bars) in a concentrated suspension for the (a) kinetic theory and (b) $B$-model. The dimensionless parameters are the same as in Fig. \ref{fig:polar-order-2d}. The $B$-model reproduces the qualitative dynamics of the kinetic theory, including the generation of $\pm1/2$ nematic defects, shown in the insets and indicated by blue squares and green triangles, respectively.}\label{fig:nematic-order-2d}
\end{figure*}

\begin{figure*}[t!]
\centering
\includegraphics[scale=0.5]{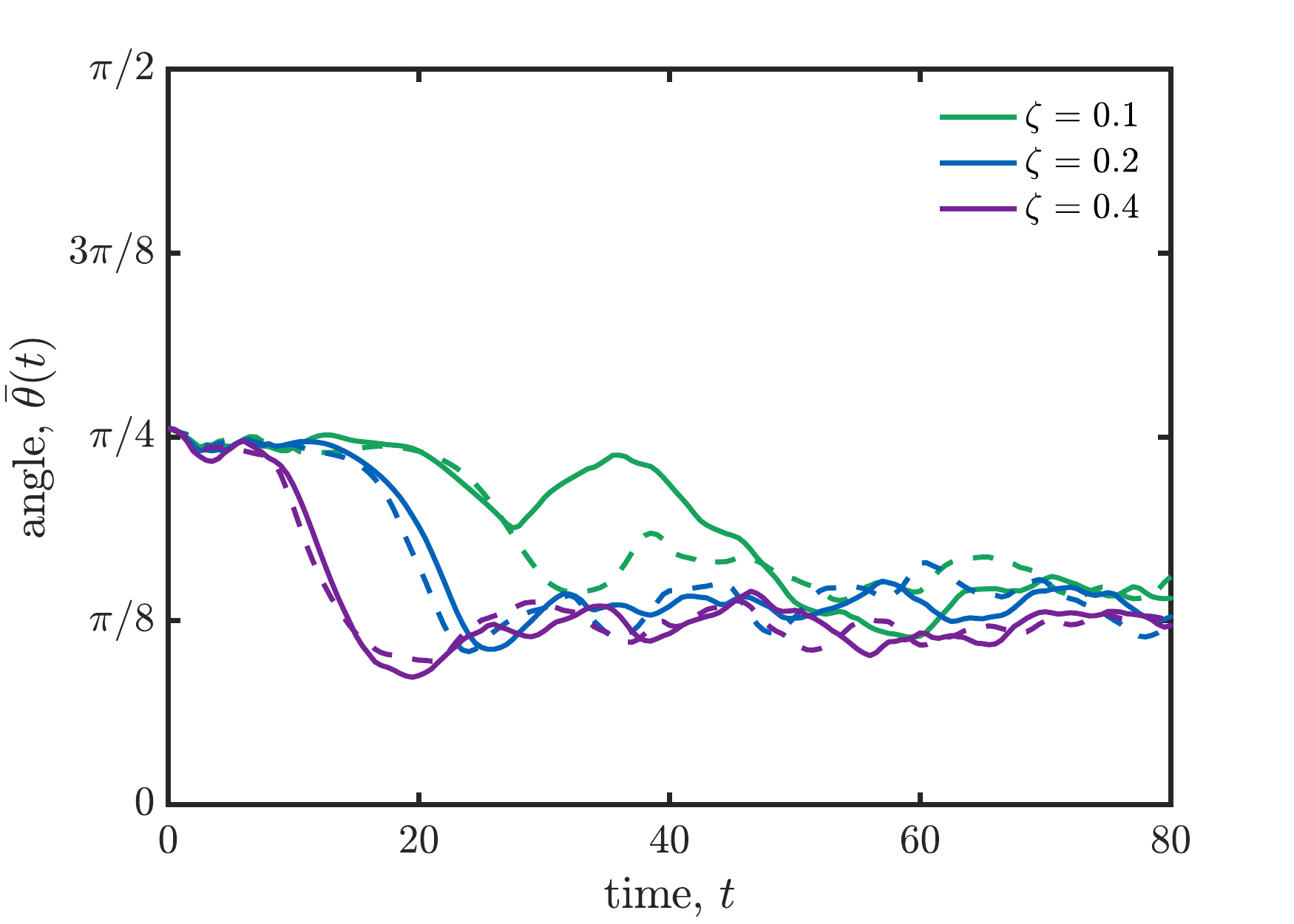}\vspace{-0.125in}
\caption{Spatially averaged angle $\bar\theta$ between the polarity vector $\n(\x,t)$ and the nematic director $\m(\x,t)$ for the kinetic theory (solid) and the $B$-model (dashed). The dimensionless parameters are $\alpha = -1$, $\beta = 0.8$, $d_T = 0.2$, and $d_R = 0.02$, and the box size is $L = 50$. For increasing $\zeta$, the $B$-model increasingly matches the kinetic theory, with the separation angle approaching $\pi/8$ over long times.}\label{fig:angle}
\end{figure*}

\begin{figure}[t!]
\centering
\includegraphics[scale=0.5]{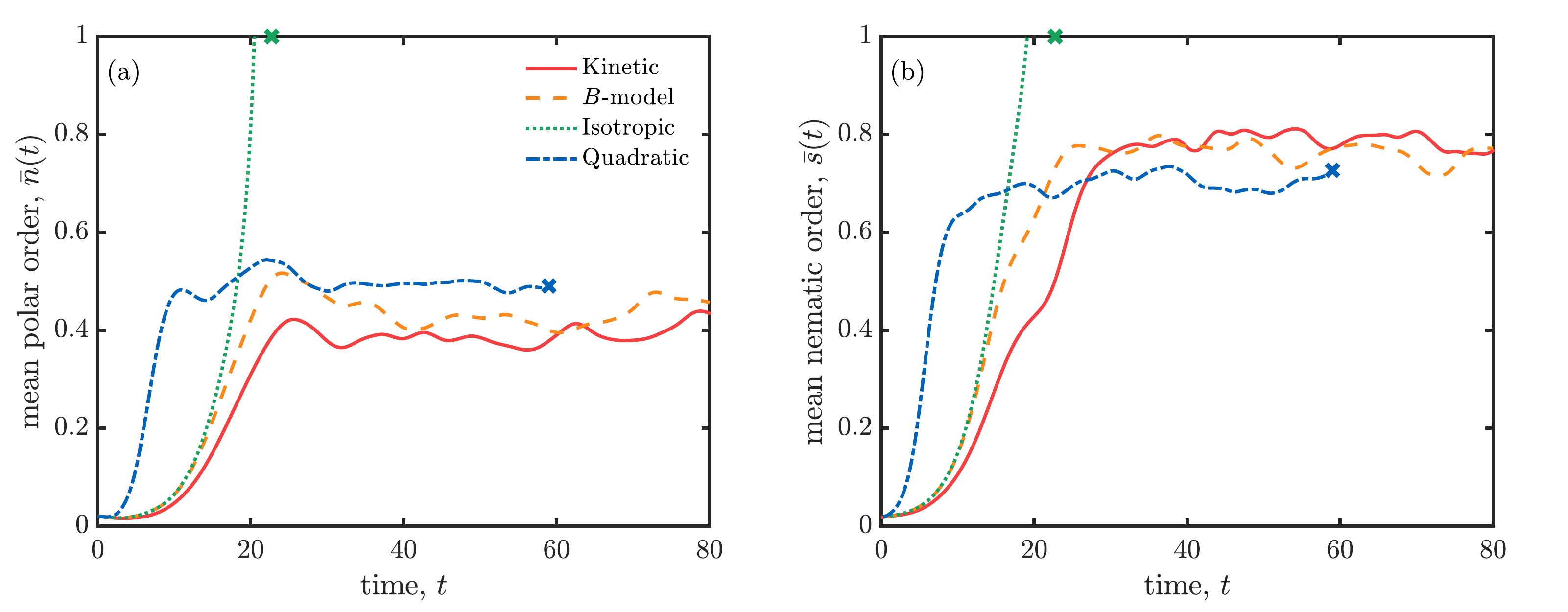}\vspace{-0.125in}
\caption{Evolution of the (a) mean polar order $\bar n(t) = \int_V |\n(\x,t)| ~ d\x$ and (b) mean nematic order $\bar s(t) = \int_V s(\x,t) ~ d\x$ in a concentrated suspension. The dimensionless parameters are $\alpha = -1$, $\beta = 0.8$, $\zeta = 0.2$, $d_T = 0.2$, and $d_R = 0.02$, and the box size is $L = 50$. For both orders, the $B$-model accurately captures the growth away from isotropy, as well as the long-time mean. When run past the initial instability, the isotropic and quadratic closures became numerically unstable with the same grid resolution regardless of the time step, with the final time denoted by $\times$.}\label{fig:order-steric}
\end{figure}

Figure \ref{fig:polar-order-2d} shows the scalar polar order parameter $n(\x,t) = |\n(\x,t)|$ along with integral curves of the polarity vector $\n(\x,t)$ at a late time from simulations of the (a) kinetic theory and (b) $B$-model. As in the dilute case, the general features and scales are consistent in both models, where we observe narrow bands of low polar order (dark blue) among broader regions of high polar order (light blue). We find numerous $\pm1$ topological defects in the polarity vector field, examples of which are shown in the insets. For the examples shown here, the $+1$ defects resemble clockwise rotating vortices, while the $-1$ defects have a saddle-point structure. We note that other types of defects are possible, including counter-clockwise rotating vortices ($-1$) and asters ($+1$) \cite{Vafa:2020}.

To characterize nematic alignment, we consider the scalar nematic order parameter
\begin{equation} s(\x,t) = \frac{d}{d-1}\bpar{\mu(\x,t) - \frac{1}{d}},\label{eq:nematic-order}\end{equation} 
where $\mu$ is the largest eigenvalue of the nematic tensor $\Q$, and the director $\m(\x,t)$, which is the eigenvector of $\Q$ corresponding to the eigenvalue $\mu$. Note that $s \approx 0$ corresponds to a state of low nematic order while $s \approx 1$ corresponds to a state of high nematic order. Figure \ref{fig:nematic-order-2d} shows a snapshot of the scalar nematic order parameter with the director field superimposed for the (a) kinetic theory and (b) $B$-model at the same time as the simulation in Fig. \ref{fig:polar-order-2d}. The two models are in good qualitative agreement, showing broad regions of high nematic alignment and numerous $\pm1/2$ nematic defects, closeups of which are shown in the insets. In both models we find that the spatial structure of the nematic field differs considerably from the polarity field. Significantly, we find polar and nematic defects do not coincide.

Differences between the polarity and director fields can be quantified by considering the separation angle $\theta(\x,t) =\arccos(\m\cdot\n/|\n|)$ between the nematic director and the polarity vector. Figure \ref{fig:angle} shows the spatially averaged angle $\bar\theta(t) = (1/V)\int_V \theta(\x,t) ~ d\x$ for both the kinetic theory and the $B$-model at several values of the alignment strength $\zeta$. For both models the polarity vector and the director are separated on average at long times by an angle of about $\pi/8$ for each value of $\zeta$ tested. This finite separation angle violates the assumption of colinearity that is implicitly used in first-moment models. Note that, because the director is unsigned, if the two vectors were uncorrelated the average angle would be $\pi/4$ for this particular branch of arccosine.

To characterize the performance of other closure models, Fig. \ref{fig:order-steric} shows the spatially averaged order parameters $\bar n(t) = (1/V) \int_V n(\x,t) ~ d\x$ and $\bar s(t) = (1/V) \int_V s(\x,t) ~ d\x$ computed from the kinetic theory, the $B$-model, and the isotropic and quadratic closures for $\zeta = 0.2$. We find polar and nematic order initially grow at comparable rates for each individual model, later fluctuating about a non-zero mean with $\bar n < \bar s$. The $B$-model gives the most accurate characterization of the kinetic theory in terms of both the initial growth rate and the long-time mean. As in the dilute case, the growth rate of the isotropic closure is similar to that of the $B$-model, while that of the quadratic closure is faster.

Even for these moderate flows, the isotropic closure became numerically unstable after the transient instability for all time steps tested (green $\times$), which we suspect is due to its inaccuracy near aligned states. This instability typically occurred when the concentration became negative, which may be a reflection of the isotropic approximation to the distribution function becoming negative at strong polar alignment. Similarly, we found the quadratic closure became numerically unstable at a later time (blue $\times$), with the instability caused by unresolved concentration gradients. These sharp gradients may be due to the artificial nonlinearities which can act to saturate smaller length scales. The $B$-model apparently does not suffer from these issues, and appears to evolve as stably as the kinetic theory.

\section{Large-scale simulations}\label{sec:big-sims}

From a computational standpoint, the coarse-grained model has a much lower cost than the kinetic theory, especially in three dimensions. In particular, the cost of the kinetic theory is $O(N_x^d N_p^{d-1})$, where $N_x$ is the number of grid points in each spatial dimension and $N_p$ is the number of grid points in each orientational degree of freedom. In contrast, the cost of the $B$-model is $O(N_x^d)$, with a modest constant that is set by the maximal degree $M$ of the Chebyshev interpolants. This makes it possible to efficiently simulate at high spatial resolutions and in aligned regimes where a large number of orientational discretization points of the distribution function are required. Here we demonstrate this capability with example large-scale two- and three-dimensional simulations. 

\begin{figure}
    \centering
    \includegraphics[scale=0.5]{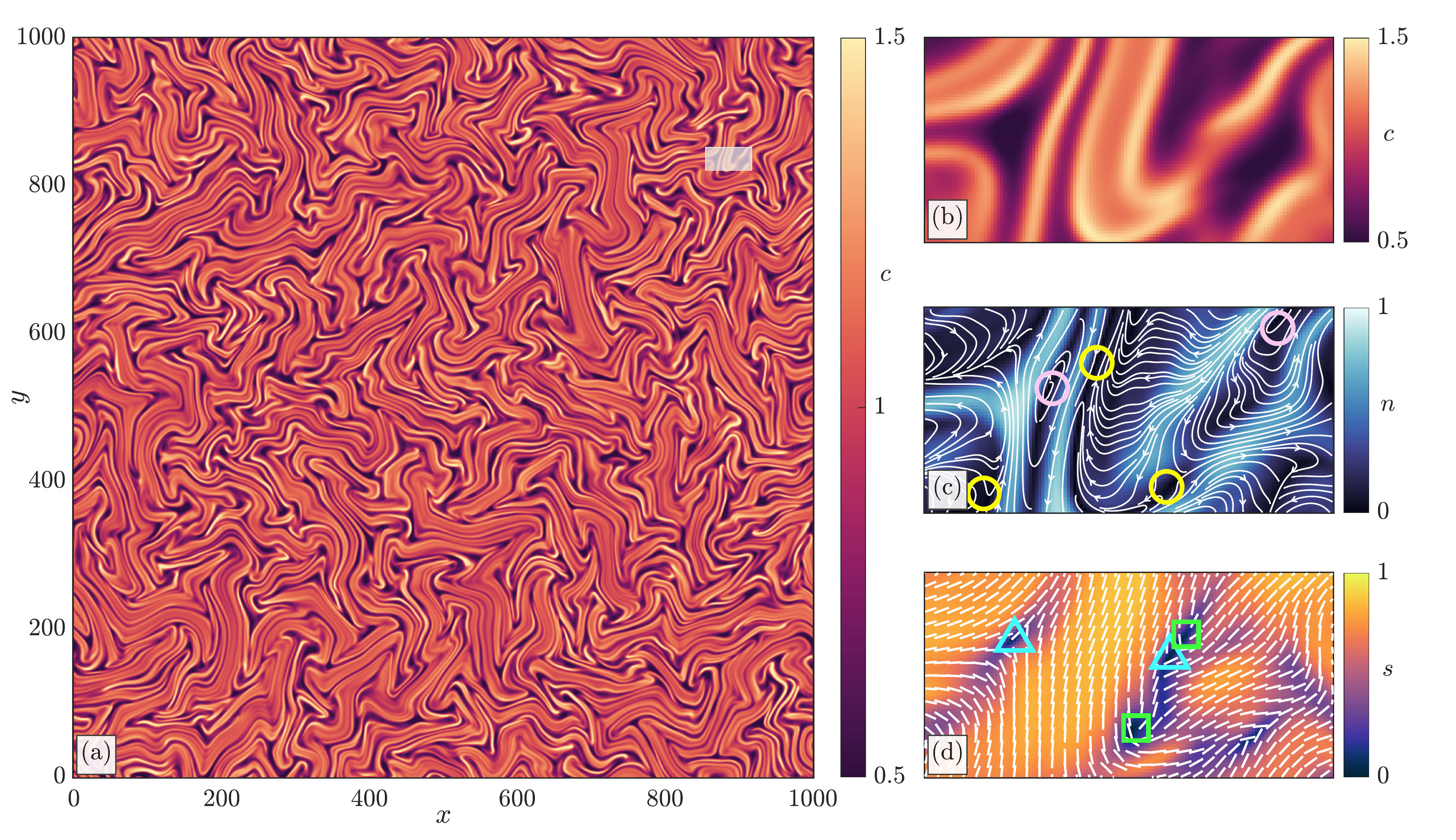}\vspace{-0.125in}
    \caption{Simulation of a two-dimensional suspension of pusher particles on a $2048^2$ grid. The dimensionless parameters are $\alpha = -1$, $\beta = \zeta = 0$, $d_T = 0.2$ and $d_R = 0.02$, and the box size is $L = 1000$. The concentration field in panel (a) shows a vast range of scales consisting of large-scale jets and fine-scale defects and striations. Closeups of the (b) concentration, (c) polar order, and (d) nematic order fields reveal a positive correlation between concentration and nematic order, and a negative correlation between concentration and polar order. $\pm 1$ topological defects are indicated by pink and yellow circles, respectively, and $\pm1/2$ defects by green squares and blue triangles, respectively.}
    \label{fig:big2d}
\end{figure}

\begin{figure}
    \centering
    \includegraphics[scale=0.5]{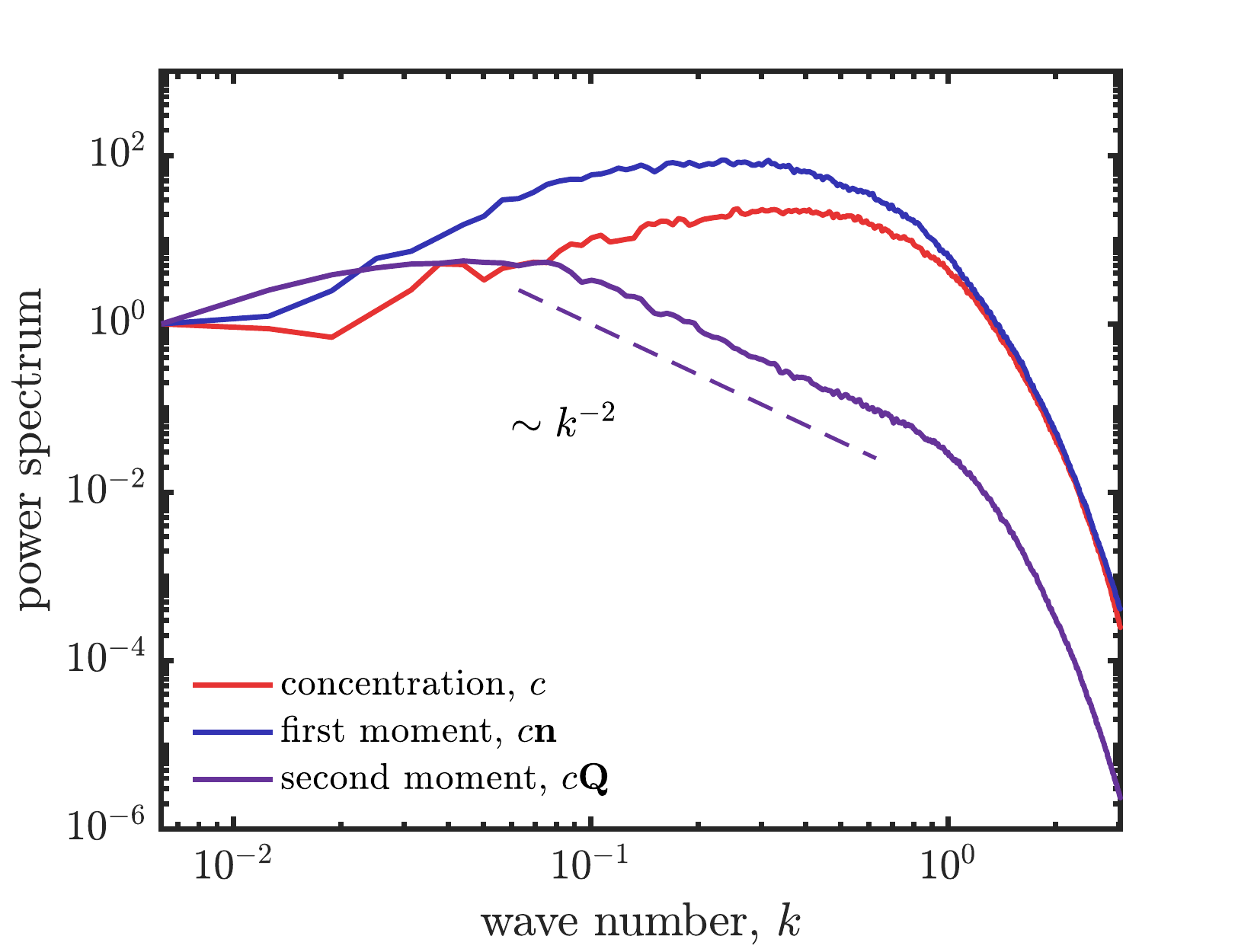}\vspace{-0.125in}
    \caption{Power spectra of the concentration $c$, first moment tensor $c\n$, and second moment tensor $c\Q$ from the two-dimensional simulation in Fig. \ref{fig:big2d}. Each field exhibits activity in length scales spanning more than two orders of magnitude. The first and second moment tensors have considerably different power spectra, with the former exhibiting a peak near $k \approx 0.1$ and the latter showing approximate $k^{-2}$ power law scaling across intermediate scales.}
    \label{fig:spectra2d}
\end{figure}

\subsection{Two dimensions}

We first consider two-dimensional suspensions with large system size $L$, for which linear stability analysis predicts a wide range of unstable wave numbers. We use the same pseudo-spectral method used in Section \ref{sec:bingham-vs-kinetic}, this time discretizing with $2048^2$ Fourier modes. The parameters are $\alpha = -1$ and $\beta = \zeta = 0$, with diffusion coefficients $d_T = 0.2$, and $d_R = 0.02$.

Figure \ref{fig:big2d}(a) shows a snapshot of the concentration field at a late time from a simulation with box size $L = 1000$. The concentration field exhibits a vast range of scales, consisting of coherent jets at large scales and striations and defect patterns at the smallest scales. Panels (b)-(d) show a zoomed-in region, indicated by the white box on panel (a), of the (b) concentration, (c) scalar polar order, and (d) scalar nematic order fields. The size of this region is $1/16 \times 1/32$ of the full box, or $128\times 64$ grid points. At these small scales, we find the polar and nematic order fields consist of numerous topological defects, with $\pm 1$ polar defects indicated by pink and yellow circles, respectively, and $\pm1/2$ defects by green squares and blue triangles, respectively. Upon inspection, we find concentration and polar order are negatively correlated, with regions of low concentration typically corresponding to regions of high polar order. On the other hand, nematic order is positively correlated with concentration, with $\pm1/2$ defects typically manifesting in similar symmetry patterns in the concentration field.

Differences between the turbulent structures of the coarse-grained fields are further identified by comparing their power spectra $[|\tilde c_\k|^2]_k$, $[|\widetilde {(c\n)}_\k|^2]_k$, and $[|\widetilde {(c\Q)}_\k|^2]_k$, where tildes denote Fourier transforms and $[\cdot]_k$ denotes an annular sum over wave numbers such that $|\k|\in[k-\Delta k,k+\Delta k]$, where $\Delta k = \pi/L$ and $k = 2\pi m/L$ for $m = 0,\ldots,N/2$. The instantaneous spectra, shown in Fig. \ref{fig:spectra2d}, identify activity in length scales ranging over nearly two orders of magnitude. These active scales extend past the marginal stability threshold of the isotropic state, which for these parameters occurs at $k = 0.5$, indicating nonlinear interactions play a strong role. Comparing each field, we find the concentration and polarity power spectra are peaked at intermediate wave numbers, reflecting characteristic nonlinear length scales. The spectra exhibit possible power law scaling from this peak to low wave numbers. In contrast, the nematic spectrum is maximal near a much longer length scale, showing near $k^{-2}$ power law scaling across intermediate wave numbers. Despite the polarity in these simulations, when taken with the Stokes equations (\ref{eq:Stokes-nd})-(\ref{eq:divu-nd}) this yields the same $k^{-4}$ power law in the velocity spectrum as that observed in active nematic turbulence \cite{Alert:2020,Martinez-Prat:2021}, suggesting nematic structure is the dominant driver of the chaotic flows.

\begin{figure}
    \centering
    \includegraphics[scale=0.5]{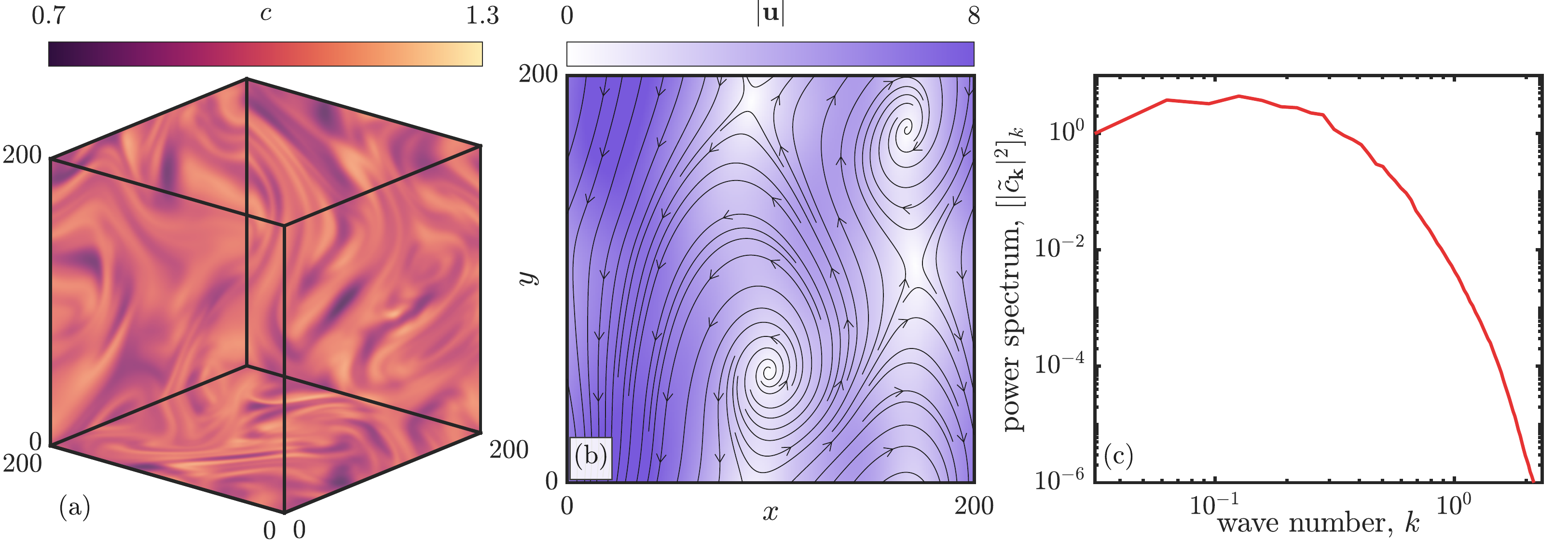}\vspace{-0.125in}
    \caption{(a) Slices of the concentration field and (b) cross section of the velocity field from a three-dimensional simulation. The grid resolution is $256^3$ and the dimensionless parameters are $\alpha = -1, \beta = \zeta = 0$, $d_T = 0.2$ and $d_R = 0.02$, and the box size is $L = 200$. (c) The power spectrum of the concentration reveals activity across more than an order of magnitude of length scales, with a characteristic scale given by the spectral peak at $k \approx 0.1$. }\label{fig:big3d}
\end{figure}

\begin{figure}
    \centering
    \includegraphics[scale=0.5]{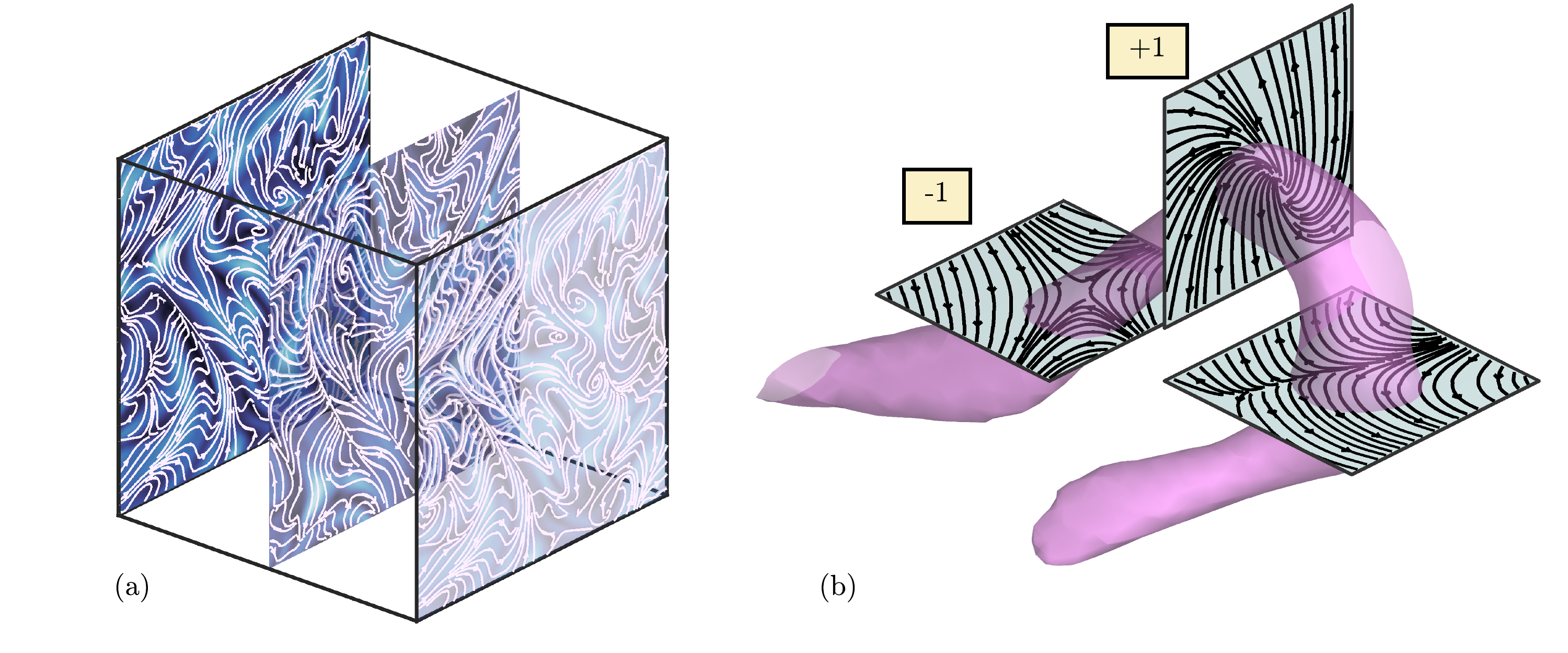}\vspace{-0.0625in}
    \caption{(a) Cross sections of the polarity field show numerous $\pm1$ planar defects with both vortex and saddle-point structures similar to those observed in two dimensions. (b) Segment of a three-dimensional polar defect. Cross sections along the defect reveal a transition between $+1$ and $-1$ defect topologies in the polarity vector field.}
    \label{fig:polar-defect}
\end{figure}

\subsection{Three dimensions}

We next illustrate an example in three-dimensions. Here we use an analogous pseudo-spectral discretization and backwards-differentiation time-stepping scheme with $256^3$ spatial Fourier modes and degree $M = 8$ Chebyshev interpolants. For context, storing the distribution function $\Psi(\x,\p,t)$ alone at this resolution assuming merely $32^2$ discretization points in orientation would require about 136GB of memory. In contrast, storing the fields $c,\n$ and $\Q$ requires less than 1GB, allowing us to run simulations at this resolution on a standard workstation.

Figure \ref{fig:big3d} shows (a) slices of the concentration field, and (b) an example cross section of the velocity field at a late time from a simulation with box size $L = 200$. The parameters here are $\alpha = -1$ and $\beta = \zeta = 0$, with diffusion coefficients $d_T = 0.2$ and $d_R = 0.02$. As in two dimensions, the concentration field consists of jets at larger scales and striations at the smallest scales, while the velocity field consists of vortices and saddle-point flows. The power spectrum of the concentration field, shown in panel (c), extends over more than an order of magnitude of wave numbers, indicating a wide range of active length scales. Similar to in two dimensions, we find the power spectrum is peaked at an intermediate wave number near $k \approx 0.1$, indicating a possible characteristic nonlinear length scale. 

Figure \ref{fig:polar-defect}(a) shows cross sections of the scalar polar order field along with integral curves of the two-dimensional projection of the polarity vector field from the same simulation in Fig. \ref{fig:big3d}. The overall structure is similar to that in two dimensions, where narrow regions of low polar order are surrounded by broader regions of high polar order. We also find numerous $\pm1$ defects with vortex and saddle-point structures, analogous to those in two dimensions. Further inspection of the polarity field shows these defects have a fundamentally three-dimensional structure. Figure \ref{fig:polar-defect}(b) shows an example portion of an isolated defect, with the overall polarity field consisting of more general filament-like defects. Integral curves of the polarity vector field along cross sections of the defect reveal a transition between $\pm1$ topologies. This is analogous to the transition between $\pm1/2$ topologies in three-dimensional nematic defects \cite{Duclos:2020}, however the precise details in the polar case are highly complex: transitions happen not only between $\pm1$ topologies, but also between the different field structures of a given topological sign. It remains for future work to characterize these structures in detail, along with the dynamics of defect nucleation and annihilation.

\section{Discussion}

We developed and analyzed a closure model and coarse-grained theory ($B$-model) for a polar active fluid. The quasi-equilibrium formulation of the closure model gives a simple prescription for constructing higher order moment closures that maintain thermodynamic consistency with the kinetic theory. Here we found that including up to the second moment tensor in the coarse-grained description was necessary for capturing salient features of the kinetic theory, including the linear instability of the isotropic, polar aligned, and nematically aligned base states, the presence of polar and nematic defects, and misalignment between the polarity vector and the nematic director. Nonlinear simulations showed the $B$-model accurately reproduces the nonequilibrium dynamics of the kinetic theory for both dilute and concentrated suspensions, especially in comparison to commonly used closure models which fail to qualitatively match the kinetic theory even in moderate flow regimes. This was verified through a range of statistics, including mean-field statistics such as polar and nematic order, and microscopic statistics such as the conformational entropy. 

To implement the closure model, we constructed Chebyshev interpolants for the closure maps in both two and three dimensions. This approach could still be improved, say by using alternative representations of the basis functions, piecewise interpolation \cite{Jiang:2021}, or analytical approximations to the full closure map. The closure model's pointwise construction means it is straightforward to parallelize and can easily be incorporated into existing codes for coarse-grained kinetic theories regardless of boundary conditions, whether in free space, in confinement \cite{Theillard:2019}, or in deformable boundaries such as active droplets \cite{Young:2021,Ruske:2021}. 

Being formulated in terms of mean-field quantities, the $B$-model establishes connections between the kinetic theory and phenomenological theories derived from symmetry considerations. In contrast to these latter theories, the evolution equations here are derived explicitly from the microscopic dynamics. This connection could be used to improve the interpretation and estimation of phenomenological parameters as well as terms that occur in the molecular free energy. 

Finally, the simulations in Section \ref{sec:big-sims} reveal many directions for future work. Tractable three-dimensional simulations allow for detailed numerical studies, which, for example, could be used to characterize the topology of three-dimensional polar defects, their dependence on particle activity, and their interplay with nematic defects \cite{Duclos:2020,Amiri:2022}. The model could also be used to study the multiscale statistics of polar active turbulence, which is largely unaddressed with full hydrodynamic theories in either two or three dimensions, all in a way that is statistically consistent with the kinetic theory.

\section*{Acknowledgements}

We thank Sebastian F\"urthauer for useful discussions in the early stages of this work, and Aleksandar Donev for providing insight into the thermodynamic interpretation. SW acknowledges support from the NSF-GRFP under Grant No. 1839302. MJS acknowledges support by the National Science Foundation under awards DMR-2004469 and CMMI-1762506.

\bibliography{closure}

\appendix

\section{Numerical computation of the closure map}\label{app:interp}

Here we construct the domain transformations between the feasible domain $(\tn,\tQ) \in \mathcal M_d$ and the hypercube $[-1,1]^{(2d-1)}$ for dimensions $d = 2,3$. We then construct Chebyshev interpolants for the maps $(\tn,\tQ)\mapsto(\tR_B,\tS_B)$ and briefly discuss their implementation.

\subsection{Two dimensions}

\begin{figure}[t!]
\centering
\includegraphics{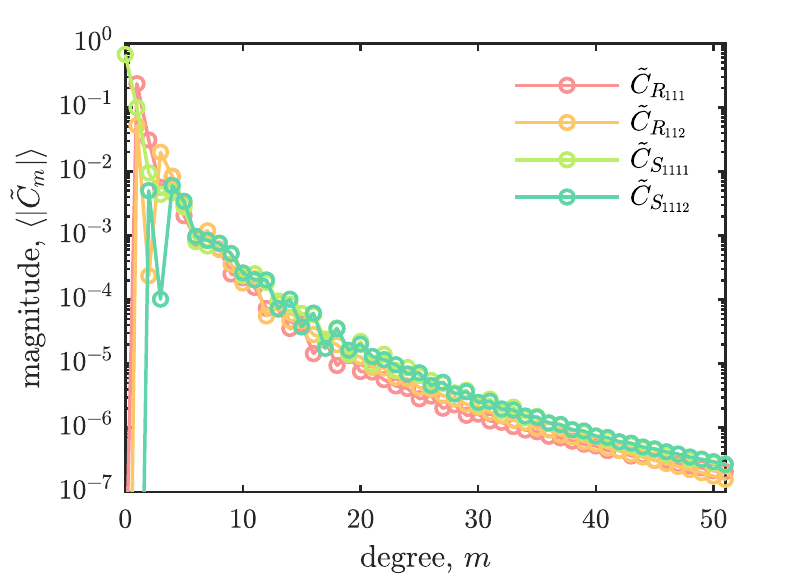}\vspace{-0.125in}
\caption{Magnitude of the Chebyshev coefficients of the two-dimensional closure maps to $\tilde R_{111},\tilde R_{112},\tilde S_{1111}$, and $\tilde S_{1112}$, averaged over degree $m$.}\label{fig:cheb-coeffs-2d}
\end{figure}

Starting first with two dimensions, the system of equations to be solved for each $(\tilde n_1,\tilde n_2,\tilde Q_{11}) \in \mathcal M_2$ is
\begin{equation}
\begin{pmatrix} 
\int_{|\p|=1} p_1 \Psi_B ~ d\p\\
\int_{|\p|=1} p_2 \Psi_B ~ d\p\\
\int_{|\p|=1} p_1^2 \Psi_B ~ d\p\\
\int_{|\p|=1} p_1p_2 \Psi_B ~ d\p
\end{pmatrix} = \begin{pmatrix} \tilde n_1 \\ \tilde n_2 \\ \tilde Q_{11} \\ 0 \end{pmatrix}.\label{eq:nonlinear-system-2d}
\end{equation}
Let $(x,y,u)\in[-1,1]^3$. We first define $\tilde Q_{11} = (3+u)/4$, which satisfies $\tilde Q_{11}\in[1/2,1]$. From the constraints $\tilde n_i^2 \leq \tilde Q_{ii} \leq 1$ and $\tilde n_1^2 + \tilde n_2^2 \leq 1$, we find the feasible range of the pair $(\tilde n_1/ \tilde Q_{11}^{1/2},\tilde n_2/ \tilde Q_{22}^{1/2})$ is the unit ball $\mathcal{B}_2 = \set{(r_1,r_2):r_1^2+r_2^2\leq 1}$. One well-conditioned invertible map from $(t_1,t_2)\in[-1,1]^2$ to $(r_1,r_2)\in \mathcal{B}_2$ is
\begin{align*}
r_1 &= t_1\bpar{1 - \frac{t_2^2}{2}}^{1/2},\\
r_2 &= t_2\bpar{1 - \frac{t_1^2}{2}}^{1/2},
\end{align*}
which yields the map $\F:(x,y,u)\mapsto(\tilde n_1,\tilde n_2,\tilde Q_{11})$,
\begin{align*}
F_1(x,y,u) &= x\bbrack{\bpar{1 - \frac{y^2}{2}}\bpar{\frac{3+u}{4}}}^{1/2},\\
F_2(x,y,u) &= y\bbrack{\bpar{1 - \frac{x^2}{2}}\bpar{\frac{1-u}{4}}}^{1/2},\\
F_3(x,y,u) &= \frac{3+u}{4}.
\end{align*}
In practice, the inverse map is evaluated with Newton's method. The nonlinear system (\ref{eq:nonlinear-system-2d}) over the square domain $[-1,1]^3$ is then
\begin{equation}
\begin{pmatrix} 
\int_{|\p|=1} p_1 \Psi_B ~ d\p\\
\int_{|\p|=1} p_2 \Psi_B ~ d\p\\
\int_{|\p|=1} p_1^2 \Psi_B ~ d\p\\
\int_{|\p|=1} p_1p_2 \Psi_B ~ d\p
\end{pmatrix} = \begin{pmatrix} F_1(x,y,u) \\ F_2(x,y,u) \\ F_3(x,y,u) \\ 0 \end{pmatrix}.\label{eq:nonlinear-system-2d_t}
\end{equation}
We assume each field can be expressed in a separable Chebyshev basis in the variables $(x,y,u)$,
\begin{align*}
\tR_B &= \sum_{\ell + m + n \leq M} \tilde\C^R_{\ell m n} T_\ell(x)T_m(y)T_n(u),\\
\tS_B &= \sum_{\ell + m + n \leq M} \tilde\C^S_{\ell m n} T_\ell(x)T_m(y)T_n(u),
\end{align*}
where $M$ is the degree of the interpolant. To compute the coefficient tensors $\tilde\C^R$ and $\tilde\C^S$, we solve Eq. (\ref{eq:nonlinear-system-2d_t}) numerically using Newton's method over a three-dimensional separable grid of Chebyshev nodes of the first kind, $t_k = \cos((2k-1)\pi/2n)$, which avoids singularities associated with the boundary of the feasible set. All of the integrals involved in the nonlinear solve are computed numerically by converting to polar coordinates $\p = (\cos\theta,\sin\theta)$, $\theta\in[0,2\pi)$, and using the trapezoidal rule in $\theta$. From these grid values we obtain the coefficient tensors from the orthogonality property of the Chebyshev polynomials applied in each variable $x,y,u$,
\begin{equation} \int_{-1}^1\frac{T_i(t),T_j(t)}{\sqrt{1-t^2}} ~ dt = \begin{cases} \pi/2 & i = j \neq 0 \\ \pi & i = j = 0 \\ 0 & i \neq j,\end{cases}\label{eq:cheb-orthogonality}\end{equation}
where the integrals are computed with Chebyshev-Gauss quadrature. Figure \ref{fig:cheb-coeffs-2d} shows the magnitude of the resulting Chebyshev coefficients up to $M = 50$, at which the remainder is $O(10^{-7})$.

\subsection{Three dimensions}

\begin{figure}[t!]
\centering
\includegraphics{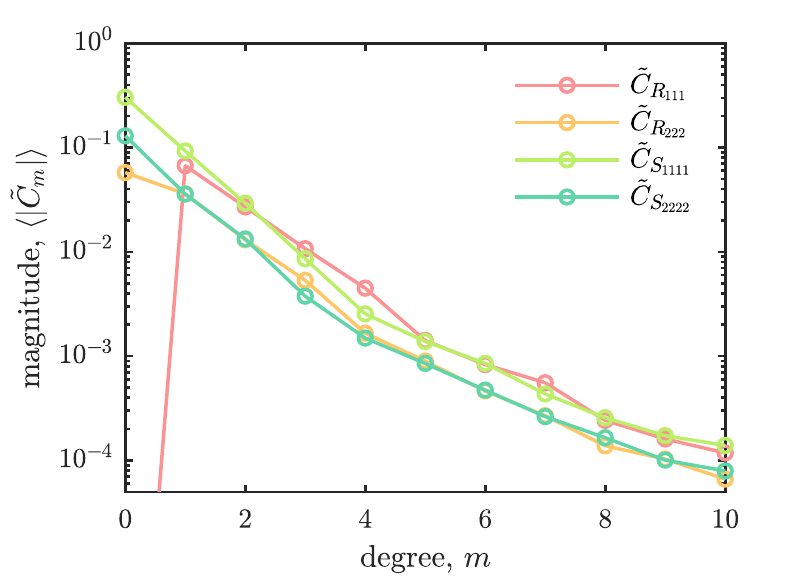}\vspace{-0.125in}
\caption{Magnitude of the Chebyshev coefficients of selected three-dimensional closure maps to $\tilde R_{111},\tilde R_{222},\tilde S_{1111}$, and $\tilde S_{2222}$, averaged over degree $m$. The coefficients of the other fields (not shown) decay at a similar rate.}\label{fig:cheb-coeffs-3d}
\end{figure}

In three dimensions the system of equations to be solved for each $(\tilde n_1,\tilde n_2,\tilde n_3,\tilde Q_{11},\tilde Q_{22})\in \mathcal M_3$ is
\begin{equation}
\begin{pmatrix} 
\int_{|\p|=1} p_1 \Psi_B ~ d\p\\
\int_{|\p|=1} p_2 \Psi_B ~ d\p\\
\int_{|\p|=1} p_3 \Psi_B ~ d\p\\
\int_{|\p|=1} p_1^2 \Psi_B ~ d\p\\
\int_{|\p|=1} p_1p_2 \Psi_B ~ d\p\\
\int_{|\p|=1} p_1p_3 \Psi_B ~ d\p\\
\int_{|\p|=1} p_2^2 \Psi_B ~ d\p\\
\int_{|\p|=1} p_2p_3 \Psi_B ~ d\p
\end{pmatrix} = \begin{pmatrix} \tilde n_1 \\ \tilde n_2 \\ \tilde n_3 \\ \tilde Q_{11} \\ 0 \\ 0 \\ \tilde Q_{22} \\ 0 \end{pmatrix}.\label{eq:nonlinear-system-3d}
\end{equation}
Let $(x,y,z,u,v)\in[-1,1]^5$. We first map the square $(u,v)\in[-1,1]^2$ to the triangular domain of eigenvalue constraints $\set{(\tilde Q_{11},\tilde Q_{22}) : 0 \leq \tilde Q_{33}\leq \tilde Q_{22}\leq \tilde Q_{11},\tilde Q_{11}+\tilde Q_{22}+\tilde Q_{33}=1}$ by composing the maps
\[ \f(u,v) = \begin{pmatrix} \frac{(1 + u)(1 + v)}{4} \\ \frac{(1 + u)(1 - v)}{4} \end{pmatrix} \]
and
\[ \g(u',v') = \begin{pmatrix} 2u'/3 + v'/6 + 1/3 \\ -u'/3 + v'/6 + 1/3 \end{pmatrix},\]
so that $\h := \g\circ\f:(u,v)\mapsto(\tilde Q_{11},\tilde Q_{22})$. Next, we make the same observation that the feasible set of $\set{\tilde n_i/\tilde Q_{ii}^{1/2}}$ is the three-dimensional unit ball $\mathcal B_3 = \set{(r_1,r_2,r_3) : r_1^2 + r_2^2 + r_3^2 \leq 1}$. One invertible map from $(t_1,t_2,t_3)\in[-1,1]^3$ to $(r_1,r_2,r_3)\in\mathcal B_3$ is
\begin{align*}
r_1 &= t_1\bpar{1 - \frac{t_2^2}{2} -  \frac{t_3^2}{2} + \frac{t_2^2t_3^2}{3}}^{1/2},\\
r_2 &= t_2\bpar{1 -  \frac{t_1^2}{2} -  \frac{t_3^2}{2} +\frac{t_1^2t_3^2}{3}}^{1/2},\\
r_3 &= t_3\bpar{1 -  \frac{t_1^2}{2} -  \frac{t_2^2}{2} +\frac{t_1^2t_2^2}{3}}^{1/2},
\end{align*}
which yields the map $\F:(x,y,z,u,v)\mapsto(\tilde n_1,\tilde n_2,\tilde n_3,\tilde Q_{11},\tilde Q_{22})$,
\begin{align*}
F_1(x,y,z,u,v) & = x\bbrack{\bpar{1 - \frac{y^2}{2} -  \frac{z^2}{2} + \frac{y^2z^2}{3}}{h_1(u,v)}}^{1/2},\\ 
F_2(x,y,z,u,v) & =y\bbrack{\bpar{1 -  \frac{x^2}{2} -  \frac{z^2}{2} +\frac{x^2z^2}{3}}{h_2(u,v)}}^{1/2},\\ 
 F_3(x,y,z,u,v) & =z\bbrack{\bpar{1 -  \frac{x^2}{2} -  \frac{y^2}{2} +\frac{x^2y^2}{3}}{h_3(u,v)}}^{1/2},\\
F_4(x,y,z,u,v) & = h_1(u,v),\\
F_5(x,y,z,u,v) & = h_2(u,v),
\end{align*}
where we've defined $h_3(u,v) = 1 - h_1(u,v) - h_2(u,v)$. As in 2D, the inverse map is evaluated with Newton's method. The system (\ref{eq:nonlinear-system-3d}) over the square domain $[-1,1]^5$ is then
\begin{equation}
\begin{pmatrix} 
\int_{|\p|=1} p_1 \Psi_B ~ d\p\\
\int_{|\p|=1} p_2 \Psi_B ~ d\p\\
\int_{|\p|=1} p_3 \Psi_B ~ d\p\\
\int_{|\p|=1} p_1^2 \Psi_B ~ d\p\\
\int_{|\p|=1} p_1p_2 \Psi_B ~ d\p\\
\int_{|\p|=1} p_1p_3 \Psi_B ~ d\p\\
\int_{|\p|=1} p_2^2 \Psi_B ~ d\p\\
\int_{|\p|=1} p_2p_3 \Psi_B ~ d\p
\end{pmatrix} = \begin{pmatrix} F_1(x,y,z,u,v) \\ F_2(x,y,z,u,v) \\ F_3(x,y,z,u,v) \\ F_4(x,y,z,u,v) \\ 0 \\ 0 \\ F_5(x,y,z,u,v) \\ 0 \end{pmatrix}.\label{eq:nonlinear-system-3d_t}
\end{equation}
As before, we assume each field can be expressed in a separable Chebyshev basis in $(x,y,z,u,v)$,
\begin{align*}
\tR_B &= \sum_{\ell + m + n + p + q\leq M} \tilde\C^R_{\ell m n p q} T_\ell(x)T_m(y)T_n(z)T_p(u)T_q(v),\\
\tS_B &= \sum_{\ell + m + n + p + q\leq M} \tilde\C^S_{\ell m n p q} T_\ell(x)T_m(y)T_n(z)T_p(u)T_q(v).
\end{align*}
To compute the coefficient tensors $\tilde \C^R$ and $\tilde\C^S$, we solve Eq. (\ref{eq:nonlinear-system-3d_t}) numerically using Newton's method over a five-dimensional separable grid of Chebyshev nodes of the first kind, along with the same orthogonality property (\ref{eq:cheb-orthogonality}). All of the integrals involved in the nonlinear solve are computed numerically by converting to spherical coordinates $\p = (\cos\phi\sin\theta,\sin\phi\sin\theta,\cos\theta)$, $\phi\in[0,2\pi)$, $\theta\in[0,\pi]$, and using the trapezoidal rule in $\phi$ and Gauss-Legendre quadrature in $\theta$. Figure \ref{fig:cheb-coeffs-3d} shows the magnitude of the Chebyshev coefficients for $\tilde R_{111},\tilde R_{222},\tilde S_{1111}$, and $\tilde S_{2222}$ up to $M = 10$, with the other fields decaying at a similar rate. For each map the remainder at $M = 10$ is $O(10^{-4})$.

\subsection{Efficient evaluation of the interpolants}

Though interpolation is more efficient than a direct solve of Eqs. (\ref{eq:nonlinear-system-2d_t}) or (\ref{eq:nonlinear-system-3d_t}), the interpolants are still relatively expensive to evaluate for large $M$, with a cost that scales as $O(M^{2d-1})$. This cost can be mitigated by noting that the components of the closure maps $(\tn,\tQ)\mapsto \tilde R_{ijk}$ and $(\tn,\tQ)\mapsto \tilde S_{ijk\ell}$ are either even or odd in each component of the rotated polarity vector $\tilde n_i$, with the symmetry determined by the even or odd number of repeated subscripts. Since the transformation $\F$ preserves the quadrant of $\tilde\n$, these symmetries are inherited by the transformed variables $x$, $y$, and $z$. Further, because the Chebyshev polynomials are also even or odd in their argument, this implies only a fourth (2D) or eighth (3D) of the terms in the coefficient tensors $\tC^R$ and $\tC^S$ are non-zero, and so these zero terms can be omitted.

\end{document}